\documentclass[twocolumn,preprintnumbers, amssymb,amsmath,aps,prd,floatfix,nofootinbib,superscriptaddress,showpacs]{revtex4-1}

\usepackage{epsfig}
\usepackage{bm}
\usepackage{amssymb}
\usepackage{amsmath} 

\usepackage{color}
\usepackage{xcolor}
\usepackage{slashed}
\usepackage[colorlinks,linkcolor=blue,anchorcolor=black,citecolor=blue]{hyperref}
\usepackage{subfigure}
\usepackage{braket}
\usepackage{physics}
\usepackage{tikz}

\usepackage{ulem}

\newcommand{\bs}[1]{\boldsymbol{#1}}

\begin{document}

\title{Nucleon Energy Correlators for the Odderon}

\author{Heikki Mäntysaari} 
\email{heikki.mantysaari@juy.fi} 
\affiliation{Department of Physics, University of Jyväskylä, P.O. Box 35, 40014 University of Jyväskylä, Finland}
\affiliation{Helsinki Institute of Physics, P.O. Box 64, 00014 University of Helsinki, Finland}

\author{Yossathorn Tawabutr} 
\email{yossathorn.j.tawabutr@jyu.fi} 
\affiliation{Department of Physics, University of Jyväskylä, P.O. Box 35, 40014 University of Jyväskylä, Finland}
\affiliation{Helsinki Institute of Physics, P.O. Box 64, 00014 University of Helsinki, Finland}

\author{Xuan-Bo Tong}
\email{xuan.bo.tong@jyu.fi} 
\affiliation{Department of Physics, University of Jyväskylä, P.O. Box 35, 40014 University of Jyväskylä, Finland}
\affiliation{Helsinki Institute of Physics, P.O. Box 64, 00014 University of Helsinki, Finland}

\begin{abstract}
We investigate the T-odd nucleon energy correlator~(NEC) at small $x$ and establish its connection with the spin-dependent odderon. Probing the T-odd NEC involves measuring a single transverse spin asymmetry~(SSA) for the energy pattern in the target fragmentation region in  deep inelastic scattering.  We find that while the inclusive energy pattern results in a vanishing SSA, a nonzero SSA emerges when restricting the measurement to charged hadrons, which are directly measured in tracking detectors. To describe this effect, we introduce the track-based NEC and evaluate its small-$x$ limit from fracture functions using the energy sum rule. The resulting SSA exhibits an opposite sign between positively and negatively charged hadrons, providing a clear experimental signature of the odderon. Furthermore, by incorporating charge weighting into the NEC, we construct a C-odd tag in the final state, eliminating potential contamination from C-even gluonic contributions. Our results demonstrate that these track-based energy-pattern observables offer a clean and sensitive probe for unveiling odderon dynamics at the future electron-ion collider.
\end{abstract}

\maketitle

\section{Introduction}
The odderon, a $C$-odd exchange, has long been predicted in high-energy hadronic scatterings~\cite{Lukaszuk:1973nt,Donnachie:1985iz}. Initially proposed to explain asymmetries between particle and antiparticle cross sections, it is now recognized as a fundamental prediction of quantum chromodynamics (QCD)~(see e.g. a review in \cite{Ewerz:2003xi}). The odderon is associated with $t$-channel colourless gluonic exchanges and plays a crucial role in high-energy QCD interactions at small-$x$~\cite{Kovchegov:2003dm,Hatta:2005as,Jeon:2005cf}. The first evidence of such exchanges was recently reported in~\cite{D0:2020tig} by comparing the measurements between the $pp$ and $p\bar p$ elastic scatterings~\cite{D0:2012erd,TOTEM:2017sdy,TOTEM:2018psk}. Despite extensive theoretical and experimental efforts in recent years~\cite{D0:2020tig,D0:2012erd,TOTEM:2017sdy,TOTEM:2018psk,Martynov:2017zjz,Lappi:2016gqe,Boer:2018vdi,Hagiwara:2020mqb,Dumitru:2021tqp,Dumitru:2022ooz,Benic:2023ybl,Benic:2024pqe,Benic:2025okp,Dumitru:2018vpr,Zhou:2013gsa,Boer:2015pni,Szymanowski:2016mbq,Hatta:2016wjz,Dong:2018wsp,Yao:2018vcg,Boussarie:2019vmk,Kovchegov:2020kxg,Kovchegov:2021iyc,Kovchegov:2022kyy,Boer:2022njw,Benic:2024fbf,Zhu:2024iwa,Benic:2024pqe}, the phenomenology of the odderon remains in its infancy, and its behavior is still poorly constrained, see~e.g.,~\cite{Hagiwara:2020mqb,Benic:2024fbf}, compared to its $C$-even counterpart, the pomeron.

Recent studies have revealed a deep connection between the odderon and proton spin physics~\cite{Kovchegov:2012ga,Zhou:2013gsa,Boer:2015pni,Szymanowski:2016mbq,Hatta:2016wjz,Dong:2018wsp,Yao:2018vcg,Boussarie:2019vmk,Kovchegov:2020kxg,Kovchegov:2021iyc,Kovchegov:2022kyy,Boer:2022njw,Benic:2024fbf,Zhu:2024iwa}. In particular, a spin-dependent odderon arises from distortions of color sources caused by the proton's transverse polarization~\cite{Zhou:2013gsa}. This spin-dependent odderon has been linked to T-odd transverse momentum distributions (TMDs), such as the well-known Sivers function~\cite{Sivers:1989cc}, identifying it as the common dynamical origin of these distributions at small-$x$~\cite{Boer:2015pni,Dong:2018wsp}.  As a consequence, the spin-dependent odderon generates a single-spin asymmetry (SSA) in hadron production during $ep^\uparrow$ collisions at small-$x$~\cite{Zhou:2013gsa,Dong:2018wsp,Zhu:2024iwa}. A key signature of the spin-dependent odderon, arising from its $C$-odd nature, is a sign flip in SSAs when the observed hadron is replaced by its corresponding anti-hadron~\cite{Dong:2018wsp,Yao:2018vcg,Zhu:2024iwa}. However, such a signal remains unobserved. The upcoming electron-ion collider (EIC) is expected to provide high-precision SSA data at small-$x$, offering a unique opportunity to probe the spin-dependent odderon~\cite{Accardi:2012qut,Aschenauer:2017jsk,AbdulKhalek:2021gbh}.

 Despite its potential, traditional hadron-level SSA observables for the odderon have several limitations. One widely studied channel is open-charm production~\cite{Zhou:2013gsa,Dong:2018wsp,Yao:2018vcg,Zhu:2024iwa}, which is sensitive to the gluonic dynamics. However, predictions for such hadron-level observables inherently depend on detailed knowledge of fragmentation functions. These functions introduce non-perturbative uncertainties from QCD hadronization when convoluted into perturbative calculations. Moreover, these SSA observables do not impose strict constraints on $C$-parity, allowing $C$-even gluonic contributions to mix with the odderon signal, leading to ambiguities in its extraction.

While the spin-dependent odderon has been well studied for the SSA in the current fragmentation region (CFR) via TMDs~\cite{Zhou:2013gsa,Boer:2015pni,Dong:2018wsp}, its role in the target fragmentation region (TFR) remains largely unexplored. In this regime, fracture functions~\cite{Trentadue:1993ka,Grazzini:1997ih,Berera:1995fj,Anselmino:2011ss}, rather than TMDs, characterize nucleon structure and describe SSAs. Factorization using fracture functions is well established to the twist-3 level~\cite{Collins:1997sr,Chen:2023wsi,Chen:2024brp,Chen:2021vby,Chai:2019ykk} and is free from Sudakov effects, which affect SSAs in the CFR. Investigating the spin-dependent odderon in the TFR is essential for a comprehensive understanding of SSAs at small-$x$, allowing us to fully exploit future EIC data in a complementary kinematical domain.

Recent advancements in energy correlator observables provide a promising framework for addressing these limitations. Unlike hadron-based observables, energy correlators measure angular distributions of total energy fluxes from final-state hadrons at collider experiments~\cite{Basham:1978bw,Basham:1977iq,Basham:1978zq}. 
These observables have been widely employed to probe QCD dynamics across a broad range of phenomena, including jet substructure~\cite{Chen:2020vvp,Li:2021zcf,Jaarsma:2022kdd,Jaarsma:2023ell,Lee:2023npz,Lee:2023tkr,Lee:2022ige,Craft:2022kdo,Komiske:2022enw,CMS:2024mlf,ALICE:2024dfl,Liu:2024lxy,Lee:2024esz,Barata:2024wsu,Alipour-fard:2024szj,Alipour-fard:2025dvp}, nuclear medium effects~\cite{Andres:2022ovj,Andres:2023xwr,Devereaux:2023vjz,Andres:2023ymw,Yang:2023dwc,Barata:2023bhh,Barata:2023zqg,Bossi:2024qho,Xing:2024yrb,Fu:2024pic,Andres:2024xvk,Barata:2025fzd,Apolinario:2025vtx} and nucleon  structure~\cite{Meng:1991da,Li:2021txc,Kang:2023big,Liu:2022wop,Chen:2024bpj,Liu:2024kqt,Liu:2022wop,Cao:2023oef,Liu:2023aqb,Li:2023gkh,Guo:2024jch,Guo:2024vpe}.
Thanks to their energy-weighted nature, energy correlator observables suppress non-perturbative effects in the final states and enhance theoretical robustness. While fully inclusive energy correlators are infared and collinear safe, this suppression remains effective even when restricting measurements to a subset of hadrons with a specific quantum numbers, such as electromagnetic charge~\cite{Chen:2020vvp,Li:2021zcf,Jaarsma:2022kdd,Jaarsma:2023ell,Lee:2023npz,Lee:2023tkr}. Notably, energy correlators on tracks (charged hadrons) have been successfully employed in precision jet substructure studies at the LHC~\cite{Komiske:2022enw,CMS:2024mlf,ALICE:2024dfl}. Moreover, as recently pointed out in \cite{Lee:2023npz,Lee:2023tkr}, a $C$-odd measurement can be achieved by incorporating charge weighting into energy correlators. Motivated by these developments, in this work, we show that measuring the energy pattern from charged hadrons provides a new probe for the odderon at the EIC.

 In particular, we apply the recently-introduced nucleon energy correlators (NECs)~\cite{Liu:2022wop}. These correlators, as an extension of the energy correlators for the nucleon structure, have been used to encode correlations between the parton structure and the energy flux in the TFR~\cite{Liu:2022wop,Chen:2024bpj,Liu:2024kqt,Cao:2023oef,Cao:2023qat,Liu:2023aqb,Li:2023gkh,Guo:2024jch,Guo:2024vpe}. Their connections to fracture functions have been elucidated in~\cite{Chen:2024bpj}. A T-odd NEC has been identified to cause SSA in the DIS energy pattern within the TFR~\cite{Liu:2022wop,Chen:2024bpj,Liu:2024kqt}. Additionally, NECs and the jet fracture functions have been shown to serve as sensitive probes of small-$x$ gluons in unpolarized targets~\cite{Liu:2023aqb,Caucal:2025qjg}, offering new insights into high-energy QCD dynamics.

The aim of this paper is to present the first application of energy correlator observables to odderon dynamics.  We investigate the energy pattern in the TFR in DIS at small-$x$ and propose the SSA  as a novel probe for the spin-dependent odderon at the EIC. While the inclusive energy pattern exhibits a vanishing SSA, we find that restricting the measurement to charged hadrons induces a nonzero SSA. We introduce the track-based NEC and establish its connection to the spin-dependent odderon. Using the energy sum rule between NECs and fracture functions, we show that the track-based NEC is directly related to the fully inclusive NEC, with the fragmentation effects entering as an overall non-perturbative factor. The track-based SSA exhibits an opposite sign between positively and negatively charged hadrons, providing a robust test for the presence of the odderon. Furthermore, by incorporating charge weighting into the NEC, we construct a $C$-odd tag on final-state events, effectively eliminating contamination from $C$-even gluonic exchanges from initial state. Our numerical results show that these track-based energy-pattern observables serve as powerful and sensitive tools for probing the spin-dependent odderon at the future EIC.

This paper is organized as follows. In Sec.~\ref{sec:observables}, we define two key energy pattern observables in our study: the track-based energy pattern and the charge patten in DIS. In Sec.~\ref{sec:DISenergypattern}, we introduce the track-based NECs and present the factorization of the DIS energy pattern in the TFR within the Bjorken limit. In Sec.~\ref{sec:NECsmallx}, we focus on the small-$x$ region and derive the connection between the track-based NEC and the spin-dependent odderon using the energy sum rule between NECs and fracture functions. In Sec.~\ref{sec:charge pattern}, we extend our analysis to the DIS charge pattern by introducing the nucleon charge correlators. In Sec.~\ref{sec:numerics}, we present our numerical prediction for the EIC. Finally, Sec.~\ref{sec:conclu} provides our conclusion.

\section{Energy-pattern observables}
\label{sec:observables}
We consider the energy pattern in deep inelastic scattering (DIS), which describes the angular distribution of hadronic energy in the final state. It was first proposed in \cite{Meng:1991da} as a generalization of the energy pattern cross section in $e^+e^-$ collisions~\cite{Basham:1977iq}. Unlike most applications of energy correlator observables studied in the current literature, the energy pattern corresponds to a one-point energy correlator, with dependence on both the polar angle ($\theta$) and the azimuthal angle ($\phi$). 

As shown in~\cite{Liu:2022wop,Chen:2024bpj,Liu:2024kqt}, an azimuthal asymmetry in the inclusive energy pattern emerges in DIS when the target nucleon is transversely polarized. In the TFR, this asymmetry, manifesting as a SSA, can be attributed to the Sivers-type quark NEC.
 In the small-$x$ regime, we will later demonstrate that the Sivers-type quark NEC originates from the spin-dependent odderon. This suggests that the SSA of the energy pattern provides a novel approach to probe the spin-dependent odderon. However, due to the $C$-odd nature of the odderon, such an access is only viable when restricting the measurement to a specific subset of hadrons, denoted as $\mathbb{S}$, to prevent SSA contributions from quark and antiquark fragmentation channels from completely canceling each other. A natural and experimentally accessible choice is to select positively or negatively charged hadrons, as these particles are directly measured in tracking detectors.

The DIS energy pattern for a subset of hadrons $\mathbb{S}$ is defined as the weighted sum over the hadronic energy $E_h$, normalized by the nucleon energy $E_N$, and integrated over the single-hadron inclusive DIS differential cross section in a given solid angle $(\theta, \phi)$~(see e.g.,~\cite{Kang:2023big,Chen:2024bpj}): 
\begin{align}
 \Sigma_{\mathbb{S}}(\theta,\phi&, \bs S_\perp)
 \notag \\  =&\sum_{h\in \mathbb{S}} \int \dd \sigma^{e +p^\uparrow\rightarrow e'+h+X} \frac{E_h}{E_N}\delta(\theta^2-\theta_h^2) \delta(\phi-\phi_h)~.
 \label{eq:energy pattern}
\end{align}
Here the sum is restricted to hadron species in $\mathbb{S}$, which in this case are e.g. hadrons with a specific charge.

Unlike the inclusive energy pattern, the charged-hadron energy pattern introduced above can receive contributions from the $C$-odd gluonic odderon, but it does not inherently isolate potential $C$-even contributions. To construct a $C$-odd tag in the final
state, we introduce the charge-weighted energy pattern in DIS: 
\begin{align}
 \Sigma_{\mathbb{Q}}(\theta,\phi&, \bs S_\perp)
 \notag \\  =&\sum_{h} \int \dd \sigma^{e +p^\uparrow\rightarrow e'+h+X} \frac{E_h Q_h}{E_N}\delta(\theta^2-\theta_h^2) \delta(\phi-\phi_h)~.
 \label{eq:charge pattern}
\end{align}  
where $Q_h$ denotes the electromagnetic charge of a final-state hadron $h$. This observable involves the multiplication of the hadronic energy $E_h$ and charge $Q_h$, providing insight into the angular distribution of the average hadronic charge. This effectively defines the charge pattern in DIS. Since neutral hadrons do not contribute to this observable, the summation $\sum_{h}$ in Eq.~\eqref{eq:charge pattern} implicitly applies only to charged hadrons. 

 We work in the Breit frame and consider a polarized nucleon with the transverse spin vector $S_\perp=|\bs S_\perp| (0,\sin \psi_S, \cos \psi_S,0)$. Here, the incoming nucleon is moving along the $z$-axis with the momentum $P $, while the virtual photon moves in the opposite direction with momentum $q^\mu=(0,0,0,-Q)$. The polar angle $\theta$ is measured relative to the nucleon beam axis, whereas the azimuthal angles $\phi$ and $\psi_S$ are measured from the plane spanned by the incoming nucleon and the electron. In this study, we focus on the energy pattern in the TFR, where the observed hadronic energy originates from the fragmentation of the target remnant. This region is kinematically characterized by $\theta \ll  1$.

For a fixed nucleon transverse polarization $S_\perp$, the Sivers-type SSA of the energy pattern can be understood as a left-right asymmetry in hadronic energy distribution. It is defined as:
\begin{align}
&A_{UT}^{\mathbb{S}}=
\notag \\ 
& ~~~2\frac{\int \dd \phi \dd \psi_s \sin \left(\phi-\psi_s\right)\big[\Sigma_{\mathbb{S}}(\theta,\phi+\pi,\bs S_\perp)-\Sigma_{\mathbb{S}}(\theta,\phi,\bs S_\perp)\big]}{\int \dd \phi \dd \psi_s\big[\Sigma_{\mathbb{S}}(\theta,\phi+\pi,\bs S_\perp)+\Sigma_{\mathbb{S}}(\theta,\phi,\bs S_\perp)\big]}~
\label{eq:SSAdef}.
\end{align}
The structure functions of the energy pattern, $\Sigma_{UU}$ and $\Sigma_{UT}$, relevant to this SSA are given as~\cite{Chen:2024bpj},
\begin{align}
&\frac{\dd \Sigma_{\mathbb{S}}(\theta, \phi, \bs S_\perp)}{\dd x_B \dd Q^2 }
=\sum_{\lambda=L,T}\frac{2\alpha_\mathrm{em}^2}{x_B Q^4}f_\lambda(y)
\notag \\ & \quad\quad\quad\times\bigg[\Sigma_{U U,\lambda}^{\mathbb{S}}+
|\bs {S}_{\perp}|\sin (\phi-\psi_S)\Sigma_{U T,\lambda}^{{\mathbb{S}},\sin (\phi-\psi_S)}\bigg]~.
\label{eq:SF}
\end{align}
Here, $f_T(y)=(1- y+y^2/2)$ and $f_L(y)=(1-y)$ denote the photon flux with the transverse (T) and longitudinal (L) polarization, respectively.
We also introduce the standard DIS kinematic variables: the photon virtuality $Q^2=-q^2$, the Bjorken variable $x_B=Q^2/(2P\cdot q)$, and the inelasticity $y=Q^2 /\left(s x_{\mathrm{B}}\right)$ with $\sqrt{s}$ as the center-of-mass energy of the $ep$ collision. 
The structure functions and SSA of the charge-weighted energy pattern are defined analogously to Eqs.~\eqref{eq:SSAdef} and~\eqref{eq:SF}.

\section{The DIS energy pattern on tracks}
\label{sec:DISenergypattern}
 \subsection{Factorization and the track-based NECs}  
 \label{sec:factorization}

We now present the factorization of the DIS energy pattern from charged hadrons in the TFR and introduce the track-based NECs. 
We are interested in the leading power contributions to the energy pattern in the TFR, where $\theta Q\ll Q$, under the Bjorken limit~($Q\gg \Lambda_{\text{QCD}}$). In this regime, the observed hadronic energy predominantly originates from the fragmentation of the target remnants after the virtual photon strikes out a parton from the target. The general factorization of the inclusive DIS energy pattern in the TFR have been demonstrated in~\cite{Chen:2024bpj,Cao:2023oef}. It follows from the observation that, while the target fragmentation dynamics remain strongly correlated with the struck parton, they are kinematically well-separated from the hard scattering process between the struck parton and the virtual photon, which is perturbatively calculable. 

The correlations between the struck parton and target fragmentation can be systematically described by NECs, $f(x,\theta,\phi)$, which encode the conditional probability of striking a parton with momentum fraction $x$ while simultaneously observing an energy flux at a solid angle $(\theta,\phi)$ in the TFR. Restricting the measurement to charged hadrons $\mathbb{S}$ does not alter the general factorization structure but requires an extension from inclusive NECs to track-based NECs. We denote the track-based NECs as $f_{\mathbb{S}}(x,\theta,\phi,\bs S_\perp)$, where we also introduced a dependence on the proton spin $\bs S_\perp$.

At the leading power, the DIS energy pattern from charged hadrons in the TFR can be factorized in terms of track-based NECs as follows:
\begin{align}
&\frac{\dd \Sigma_{\mathbb{S}}(\theta, \phi, \bs S_\perp)}{\dd x_B \dd Q^2 }
=\frac{2\alpha_\mathrm{em}^2 }{ Q^4}\sum_{\lambda=L,T}f_\lambda(y)
\notag \\ 
&\quad\quad\times\sum_{a}\int_{x_B}^1 \frac{\dd x}{x} {\cal H}_{a,\lambda}\Big(\frac{x_B}{x},\frac{Q}{\mu}\Big) 
f^a_{\mathbb{S}}(x,\theta,\phi, \bs S_\perp,\mu)~.
\label{eq:factorization}
\end{align} 
Here, ${\cal H}_{a,\lambda}$ represents the hard partonic cross-section for the scattering between a struck parton $a$ and the virtual photon with the polarization $\lambda=L,T$. This factorization follows collinear factorization since no energy flux is observed from the current fragmentation. As a result, the transverse momentum of the struck parton remains unresolved in the hard scattering and can be neglected in the Bjorken limit. Additionally, soft gluons decouple and cancel due to color coherence in the inclusive sum over unobserved hadrons, as in inclusive DIS~\cite{Collins:1997sr}. Consequently, the hard scattering kernels ${\cal H}_{a,L/T}$ are identical to those in inclusive DIS to all orders in $\alpha_s$. The NEC $f^a_{\mathbb{S}}$ follows the standard DGLAP evolution equation, with the factorization scale denoted as $\mu$ in Eq.~\eqref{eq:factorization}.

The track-based NECs can be defined analogously to inclusive NECs~\cite{Liu:2022wop,Chen:2024bpj,Liu:2024kqt}. We follow the notation in \cite{Chen:2024bpj}. The leading-twist quark contributions relevant to the SSA are given by the following correlation matrix:
\begin{align}
f^q_{\mathbb{S}}&(x,\theta,\phi, \bs S_\perp)
= \int  \frac{\dd y^-}{4\pi}e^{-ix P^+y^-} 
\notag \\
&\times
\langle PS_\perp|\bar\psi(y){\cal L}^\dagger(y)\gamma^+{\cal E}_{\mathbb{S}}(\theta,\phi) {\cal L}(0)\psi(0)|PS_\perp\rangle
\notag \\
=&f^q_{1,\mathbb{S}}\left(x, \theta\right)+\frac{\epsilon^{\perp}_{\mu \nu} \bs S_\perp^{\mu}\bs {n}_{t}^\nu}{|\bs n_t|} f^{t,q}_{1T,\mathbb{S}}\left(x,\theta\right).
\label{eq:NEEC_def}
\end{align}
Here we have used the light-cone variables for a vector as $a^\mu = (a^+,a^-, \vec a_\perp) = \bigl((a^0+a^3)/{\sqrt{2}}, (a^0-a^3)/{\sqrt{2}}, a^1, a^2 \bigr)$, $\bs n_t^\mu= \sin\theta (\cos\phi,\sin\phi)$ denotes the azimuthal vector of the observed energy flux, and the transverse antisymmetric tensor is defined as $\epsilon_\perp^{\mu\nu} = \epsilon^{\mu\nu-+}$, with $\epsilon^{0123}=1$. The light-like gauge link, ${\cal L}(\xi)$, is process dependent and extends to either $+\infty$ or $-\infty$ due to final-state or initial-state interactions. For DIS processes, we take the gauge link pointing to $+\infty$:
\begin{align}
{\cal L}(\eta^-,\eta_\perp)=\text{P}~\text{exp}\left[- i g_s\int_{\eta^-}^{+\infty} \dd \xi^- A^{+}(\xi^-,\eta_\perp)\right]~,
\label{eq:L}
\end{align}
where the gluon field $A^\mu$ is in the fundamental representation.

The energy flux measurement within the subset $\mathbb{S}$ is captured by the operator ${\cal E}_{\mathbb{S}}(\theta,\phi)$ in Eq.~\eqref{eq:NEEC_def}. This operator can be expressed as the hadronic number operator weighted by energy~\cite{Chen:2024bpj,Sveshnikov:1995vi,Bauer:2008dt}:
 \begin{align} {\cal E}_{\mathbb{S}}(\theta,\phi)=\sum_{h\in \mathbb{S} } \int \frac{\dd P_h^+ \dd^2\bs P_{\perp} }{2P_h^+(2\pi)^3}\frac{E_h}{E_N}\delta(\theta^2-\theta^2_h)\delta(\phi-\phi_h)a^\dagger_h a_h, \label{eq:sum1} \end{align} where $a^\dagger_h a_h$ is the number operator of a hadron $h\in\mathbb{S} $ with momentum $P_h$.  The inclusive case can be recovered by summing over all hadron species.

The functions $f_{1,\mathbb{S}}$ and $f^{t}_{T,\mathbb{S}}$ in Eq.~\eqref{eq:NEEC_def} describe the unpolarized quark distributions in an unpolarized and transversely polarized nucleon, respectively~\cite{Chen:2024bpj}. In particular, $f^{t}_{T,\mathbb{S}}$, known as the Sivers-type NEC, generates the SSA given in Eq.~\eqref{eq:SSAdef}. This function encodes a T-odd effect, as it changes sign under the $PT$ transformation and vanishes if the gauge link in Eq.~\eqref{eq:NEEC_def} accounting for final-state interactions is neglected. 

The factorization formula in Eq.~\eqref{eq:factorization} shows that the two quark NECs $f^q_{1,\mathbb{S}}$ and $f^{t,q}_{T,\mathbb{S}}$ share the same hard coefficients and evolve according to the same DGLAP equation. In general, there are contributions from gluon NECs, which can be defined analogously to the quark NECs. However, in this work, we focus on the leading order~(LO) in $\alpha_s$, where only the quark and anti-quark channels contribute\footnote{In fact, we safely neglect the gluon NECs even at higher orders of $\alpha_s$, because the gluon NECs do not give rise to the spin-dependent odderon at small-$x$. We will demonstrate this in a separate work.}. Similar to inclusive DIS, the contributions from longitudinally polarized photons to the energy pattern are absented at LO in $\alpha_s$~\cite{Chen:2024bpj}. Therefore, the relevant hard coefficients are ${\cal H}_{a,T}(x)=e_a^2\delta(x)+\mathcal{O}(\alpha_s)$ for $a=q,\bar q$. The associated energy pattern structure functions in Eq.~\eqref{eq:SF} can be expressed as
\begin{align}
\Sigma_{UU,T}^{\mathbb{S}}=&\frac{2\alpha_\mathrm{em}^2 }{ Q^4}f_T(y)\sum_{a=q,\bar q} e_a^2 f^a_{1,\mathbb{S}}(x_B,\theta)~,
 \\ 
    \Sigma_{U T,T}^{\mathbb{S}}=&\frac{2\alpha_\mathrm{em}^2 }{ Q^4}f_T(y)\sum_{a=q,\bar q} e_a^2 f^{t,a}_{1T,\mathbb{S}}(x_B,\theta)~.    
\end{align}
With the above results, the SSA defined in Eq.~\eqref{eq:SSAdef} can be written as the ratio of the Sivers-type and unpolarzied quark track-based NECs:
\begin{align} &A_{UT}^{\mathbb{S}}=\frac{ \Sigma_{U T,T}^{\mathbb{S}}}{\Sigma_{UU,T}^{\mathbb{S}}}= \frac{\sum_{a=q,\bar q} e_a^2 f^{t,a}_{1T,\mathbb{S}}(x_B,\theta) }{\sum_{a=q,\bar q} e_a^2 f^a_{1,\mathbb{S}}(x_B,\theta)}~. 
\label{eq:SSANEC}
\end{align}
It is noted that only the flavor-singlet contributions are relevant in the SSA.

 \subsection{Connection to Fracture Functions}

If the energy operator ${\cal E}_{\mathbb{S}}(\theta,\phi)$ in the correlation matrix in Eq.~(\ref{eq:NEEC_def}) is replaced by the number operator $a^\dagger_h a_h$ of a given hadron $h$, one obtains the associated quark fracture functions $u_1$ and $u_{1T}^h$ for measuring the hadron $h$:
\begin{align}
{ u}^q_h(x, &\xi_h, \bs P_{h\perp}, \bs S_\perp)
\notag \\ 
=&\int  \frac{\dd y^-}{4\xi_h(2\pi)^4}e^{-ix P^+y^-} \langle PS_\perp|\bar\psi(y){\cal L}^\dagger(y)\gamma^+ |P_h,X\rangle
\notag \\
&\times\sum_X 
\langle P_h,X|  {\cal L}(0)\psi(0)|PS_\perp\rangle
\notag \\
=&{u}_{1}^q\left(x,\xi_h, \bs P_{h\perp}^2\right)+\frac{\epsilon^{\perp}_{\mu \nu} \bs S_\perp^{\mu}\bs {P}_{h\perp}^\nu}{M} {u}^{h,q}_{1T}(x,\xi_h, \bs P_{h\perp}^2)~,
\label{eq:Frac_def}
\end{align}
where we have used the identity $a^\dagger_h a_h=\sum_X |P_h,X\rangle\langle P_h,X| $ and included a normalization factor $1/(2\xi_h(2\pi)^3)$ for convention~\cite{Chen:2023wsi}. The variable $\xi_h = P_h^+/P^+$ represents the fraction of the target momentum carried by the observed hadron. Fracture functions describe the parton distributions inside the target under the condition that the target remnants fragment into a specific hadron $h$ with momentum $P_h$. These functions play a crucial role in understanding the hadronization process in the TFR and have been shown to generate various azimuthal correlations in single-hadron and double-hadron inclusive production in DIS~(see e.g.,~\cite{Anselmino:2011ss,Chen:2023wsi,Chen:2024brp,Chen:2021vby,Chai:2019ykk,Anselmino:2011bb,Anselmino:2011vkz,CLAS:2022sqt,Guo:2023uis,Zhao:2024usu})

Using the relation in Eq.~\eqref{eq:sum1}, an energy sum rule has been established in~\cite{Chen:2024bpj}, demonstrating that fracture functions serve as the parent functions of the associated inclusive NECs, with a direct one-to-one correspondence between them. This sum rule also works for the track-based NEC. Specifically, the track-based NEC $f^q_\mathbb{S}$ defined in Eq.~\eqref{eq:NEEC_def} can be expressed in terms of the fracture function defined in Eq.~\eqref{eq:Frac_def} as: 
\begin{align}
 f^q_{\mathbb{S}}&(x,\theta,\phi, \bs S_\perp)
 \notag \\
 =&\sum_{h\in \mathbb{S} } \int^{1-x}_0 \dd\xi_h \int \dd[2]\bs P_{\perp} \xi_h\delta(\theta^2-\theta^2_h)\delta(\phi-\phi_h)
  \notag \\
 &~~~~\times u^q_h(x,\xi_h, \bs P_{h\perp}, \bs S_\perp)
\notag \\
=&
 \sum_{h\in \mathbb{S} }  \int_0^{1-x} \dd\xi_h~ \xi_h\frac{\bs P_{h\perp}^2}{2\theta^2} u^q_h(x,\xi_h, \bs P_{h\perp}, \bs S_\perp)
\Big\vert_{\substack{ \bs P_{h\perp} = \frac{ \xi_h P^+ }{\sqrt{2}}}\bs n_t}\,.
\label{eq:connection}
 \end{align}
 This sum rule is essentially a manifestation of energy conservation: the total hadronic energy must equal the sum of the energies of all hadron species under consideration.  
 As we will show later, this sum rule is useful in evaluating the track-based NEC at small-$x$. Furthermore, employing this sum rule, the factorization formula for the track-based energy pattern can be systematically derived from the well-established factorization in SIDIS, following the procedure outlined in \cite{Chen:2024bpj}.

\begin{figure}[t]
\centering
\includegraphics[scale=0.70]{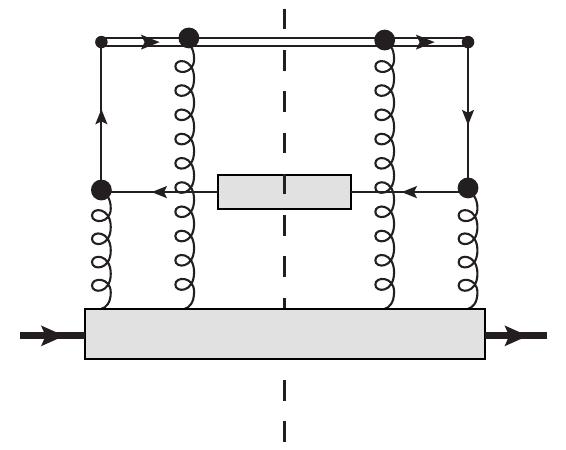}
\caption{The LO contribution of the small-$x$ quark NECs. The double line represents the light-cone gauge link in Eq.~(\ref{eq:NEEC_def}) for the DIS process. The quark-gluon vertex and the eikonal-gluon vertex here represents the interaction from the shock wave, which contain multiple-gluon interactions with small-$x$ gluons from the target $A$.}
\label{fig:amp}
\end{figure}

\section{NECs at small-x: Connection to the spin-dependent odderon }
\label{sec:NECsmallx}
In the previous section, we demonstrated that the SSA of the energy pattern is determined by the flavor-singlet contribution of the Sivers quark NEC: $\sum_{a=q,\bar q} e_a^2 f^{t,a}_{1T,\mathbb{S}}(x_B,\theta)$. In this section, we focus on the small-$x_B$ region and establish the connection between the Sivers quark NEC and the spin-dependent odderon. We first analyze the inclusive case, $\mathbb{S}=$all hadrons, and show that the flavor-singlet component of the inclusive Sivers quark NEC vanishes at small-$x$ due to the $C$-odd nature of the odderon. To access the spin-dependent odderon, we then evaluate the track-based NEC, which restricts the measurement to a specific subset of charged hadrons, $\mathbb{S}$. Using the energy sum rule, we derive the small-$x$ behavior of the track-based NEC from fracture functions and demonstrate that it leads to a nonzero flavor-singlet Sivers NEC at small-$x$, providing a novel probe for the spin-dependent odderon.

\subsection{The inclusive NEC at small-$x$}
 
In the TFR, the polar angle of the NECs remains relatively small, $\theta Q \ll Q$. When $\theta Q \sim \Lambda_{\text{QCD}}$, the correlations between the initial parton (or target) and the energy flow are generally non-perturbative. However, in the small-$x$ regime, these correlations can become perturbatively accessible due to the emergence of the saturation momentum $Q_s$, where $Q_s \gg \Lambda_{\text{QCD}}$. To probe these small-$x$ effects, one can position the detectors at an angle satisfying $Q_s  \lesssim \theta Q \ll Q$. 
As we will argue in Sec.~\ref{sec:numerics}, this part of the phase space will be covered by the EIC detectors.

 In this region, the quark NECs can be computed within the color glass condensate (CGC) formalism~\cite{Gelis:2010nm,Iancu:2003xm} using perturbation theory. At leading logarithmic accuracy, the contributions arise from the perturbative diagrams shown in Fig.~\ref{fig:amp}, where a quark-antiquark pair is generated from small-$x$ gluons in the target, with a characteristic transverse momentum $|\bs k_{\perp}| \lesssim Q_s$. For the quark NEC, the quark becomes the parton initiating the hard scattering~(e.g., with the virtual photon), while the antiquark undergoes fragmentation, producing hadrons that contribute to the measured energy flux. The case for the antiquark NEC follows a similar mechanism, but with the roles reversed: the energy flux is instead initiated by the quark. 

 For the inclusive NECs, where $\mathbb{S} = \text{all hadrons}$, the production of the energy flux is perturbatively calculable to all orders of $\alpha_s$. This is because the energy weighting suppresses soft singularities, while the summation over all final-state hadrons cancels collinear singularities. Thus, one can safely perform the calculation directly at the parton level, without considering the parton-hadron transitions. In this case, the computation of small-$x$ quark NECs follows similar techniques as used for quark TMDs, as detailed in~\cite{Marquet:2009ca, Xiao:2017yya}. Below, we summarize the key results for the inclusive NEC, where we suppress the subscript $\mathbb{S} = \text{all hadrons}$.

  We find that the small-$x$ quark NEC can be expressed in terms of the dipole gluon distributions $ {\cal F}_{x_g}$:
  \begin{align}
f^q(x,&\theta,\phi, \bs S_\perp)
=\frac{N_c}{\theta^2 (2\pi)^4 }
\int_0^{1-x}\dd \xi
\notag \\
& 
\times \int \dd[2]\bs k_{g\perp}\bs k_\perp^2{\cal H}_q(\xi,\bs k_\perp,\bs k_{g\perp})
 {\cal F}_{x_g}(\bs k_{g\perp},\bs S_\perp)~,
\label{eq:smallNEEC}
\end{align}
    where $\xi P^+$ and $\bs k_{\perp}\equiv \frac{\xi P ^+}{\sqrt{2}}\bs n_t$ represent the longitudinal and transverse momenta of the outgoing anti-quark in the TFR, respectively. The perturbative coefficient ${\cal T}_q$ has the following form:
    \begin{align}
    {\cal H}_q(\xi,\bs k_\perp,\bs k_{g\perp})=\frac{\bs k_\perp^2}{\xi} \Bigg[ 
    \frac{\bs  k_{g\perp} +\bs  k_{\perp} }{\epsilon_f^2+(\bs  k_{g\perp} +\bs  k_{\perp})^2}
    -\frac{\bs  k_{\perp} }{\epsilon_f^2+\bs k_{\perp}^2}
    \Bigg]^2, 
    \label{eq:Afunction}
    \end{align}  
    where $\epsilon_f^2\equiv(x/\xi) \bs k^2_{ \perp}$.
    
     The dipole gluon distribution $ {\cal F}_{x_g}$ is the Fourier transform of the quark dipole S-matrix, defined as: 
    \begin{align}
     {\cal F}_{x_g}&(\bs k_{g\perp},\bs S_\perp)=\int \frac{\dd[2] \bs b_{\perp} \dd[2] \bs r_{\perp}}{(2 \pi)^2} e^{-i \bs k_{g\perp} \cdot\bs  r_{\perp}} 
     \notag \\
     &\times \frac{1}{N_c}\left\langle \text{tr} \big[ U(\bs b_{\perp}+\frac{\bs r_{\perp}}{2}) U^{\dagger}(\bs b_{\perp}-\frac{\bs r_{\perp}}{2})\big]\right\rangle_{x_g}~,
    \label{eq:quarkdipole}
    \end{align}
Here, $U(\bs{x}_{\perp})= {\cal L}(-\infty,\bs{x}_{\perp})$ represents the Wilson line, extending from $x^{-}=-\infty$ to $x^{-}=+\infty$ at the transverse position $\bs{x}_{\perp}$, and $\langle \cdot \rangle_{x_g}$ represent an average over the target configurations. In coordinate space, the quark dipole S-matrix describes the multiple interactions between small-$x$ gluons in the target and a fast-moving quark-antiquark pair, positioned at $\bs b_{\perp}+\bs r_{\perp}/2$ and $\bs b_{\perp}-\bs r_{\perp}/2$, respectively. Here, $\bs r_\perp$ represents the dipole orientation, while $\bs b_\perp$ denotes the impact parameter. Under charge conjugation ($C$-parity transformation), the roles of the quark and antiquark are interchanged, which corresponds to reversing the dipole orientation, $\bs r_\perp \rightarrow -\bs r_\perp$. In momentum space, this transformation translates to reversing the direction of the transverse momentum transfer, $\bs k_{g\perp}$, from the target.
    
 The dipole gluon distribution consists of two distinct terms with different $C$-parity properties:
    \begin{align}
      {\cal F}_{x_g}(\bs k_{g\perp},\bs S_\perp)=  F_{x_g}( \bs k_{g\perp}^2)
    +\frac{\epsilon_{\perp}^{i j} \bs S_{\perp i} \bs k_{g\perp j}}{M}  O_{1 T,x_g}^{\perp}( \bs  k_{g\perp}^2)~.
    \label{eq:quarkdipole2}
    \end{align}
 The first term, $F_{x_g}( \bs k_{g\perp}^2)$, represents the standard dipole pomeron distribution and is $C$-even, since it remains unchanged under $\bs k_{g\perp}\rightarrow -\bs k_{g\perp}$. The second term, $O_{1T,x_g}^{\perp}(\bs k_{g\perp}^2)$, known as the spin-dependent odderon~\cite{Zhou:2013gsa}, encodes a C-odd correlation between the transverse momentum transfer from the target and the target’s transverse spin. Due to its linear dependence on $\bs k_{g\perp}$, this term changes sign under charge conjugation, confirming its $C$-odd nature. In the dilute limit, the spin-dependent odderon is associated with tri-gluon correlations  $\sim d^{abc} A^{a,+} A^{b,+} A^{c,+}$, where $d^{abc}$ is the color symmetric structure constant (see Eq.~\eqref{eq:tri1} in the Appendix~\ref{sec:appendix}) .

    Comparing the results in Eqs.~\eqref{eq:smallNEEC} and \eqref{eq:quarkdipole2}, with the definition of the track based NEC in Eq.~\eqref{eq:NEEC_def}, we observe that the unpolarized quark NEC $f^q$ is non-zero due to the pomeron distribution:
    \begin{align}
     f^{q}_1(x, \theta)=&\frac{ N_c}{\theta^2(2\pi)^4 }
    \int_0^{1-x}\dd \xi 
    \notag \\ 
    & \times 
    \int \dd[2]\bs k_{g\perp}\bs k_\perp^2{\cal H}_q(\xi,\bs k_\perp,\bs k_{g\perp}) F_{x_g}( \bs k_{g\perp}^2)
    ~.
    \label{eq:qNEEC_P}
    \end{align}
    This result is equivalent to the expression derived earlier in~\cite{Liu:2023aqb}. Since the pomeron is $C$-even, the unpolarized antiquark NEC $f^{\bar q}_1$ share the same expression with the corresponding quark NEC, $f^{ q}_1$.
    
    More interestingly, we find that the quark Sivers NEC originates from the spin-dependent odderon: 
    \begin{align}
    f^{t,q}_{1T}&(x, \theta)=
    \frac{ N_c}{\theta^2(2\pi)^4 }  \int_0^{1-x}\dd \xi 
    \notag \\ 
    & \times \int \dd^2 \bs k_{g\perp} \bs k_\perp^2 {\cal H}_q(\xi,\bs k_\perp,\bs k_{g\perp})  \frac{\bs k_\perp\cdot \bs k_{g\perp }}{|\bs k_\perp|} O_{1 T,x_g}^{\perp}( \bs  k_{g\perp}^2) ~.
    \label{eq:qNEEC_O}
    \end{align} 
    Since the Sivers NEC governs the SSA of the DIS energy pattern, the direct connection established in Eq.~\eqref{eq:qNEEC_O} implies a novel method to probe the spin-dependent odderon through the energy flux measurements in the TFR.

     However, the practical realization is complicated by the $C$-odd nature of the odderon, because the anti-quark Sivers NEC exhibits a opposite sign compared to the quark one: 
     \begin{align}
         f^{t,\bar q}_{1T}&(x, \theta)=- f^{t, q}_{1T}(x, \theta)~. 
         \label{eq:fqop}
     \end{align}
   This sign difference can be directly verified through a straightforward calculation. First, we note that the perturbative coefficient for the anti-quark NEC is related to that of the quark NEC by: \begin{align}
{\cal H}_{\bar q}(\bs k_\perp,\bs k_{g\perp})={\cal H}_q(\bs k_\perp,-\bs k_{g\perp})~.
\end{align}
Then, performing a variable transformation $\bs k_{g\perp} \rightarrow -\bs k_{g\perp}$ in the integral, we see that the anti-quark Sivers NEC acquires an overall sign relative to the quark Sivers NEC. This arises from the linear dependence of the spin-dependent odderon term,
\begin{align} \frac{\bs k_\perp\cdot \bs k_{g\perp} }{|\bs k_\perp|} O_{1T,x_g}^{\perp}(\bs k_{g\perp}^2)~, \end{align} which flips sign under $\bs k_{g\perp} \rightarrow -\bs k_{g\perp}$ due to the $C$-oddness.

     Consequently, for the inclusive measurement, $\mathbb{S}=$all hadrons, the flavor-singlet contribution to the Sivers NECs vanishes at small-$x$: 
    \begin{align}
    \sum_{a=q,\bar q} e_a^2 f^{t,a}_{1T}(x, \theta)=0~.
    \label{eq:inclu1}
    \end{align}
This implies that the inclusive DIS energy pattern does not provide access to the spin-dependent odderon at small-$x$, since the Sivers NEC contributes solely through the flavor-singlet channel in such a process (see Eq.~\eqref{eq:SSANEC}). This issue lies in the inclusive summation over final-state hadron species, where the contributions from quark and antiquark fragmentation to the energy flux are equal.  Due to the sign difference induced by the odderon dynamics in the initial state, these contributions cancel exactly, leading to the null result in Eq.~\eqref{eq:inclu1}.

    \subsection{NECs for a subset of hadrons $\mathbb{S}$ at small-$x$}

To measure the spin-dependent odderon, it is essential to obtain a non-vanishing contribution to the flavor-singlet Sivers NECs. This can be achieved by restricting the measurement to the energy flux of a specific subset of hadrons, $\mathbb{S}$. By selecting an appropriate subset, the measured energy flux can exhibit different sensitivities to the quark and antiquark fragmentation channels, preventing complete cancellation and ensuring that the associated flavor-singlet contribution remains nonzero.  

A natural and experimentally accessible choice is to select either positively or negatively charged hadrons. Due to the charge conservation, quark and antiquark fragmentation contribute asymmetrically to these charged particles. Moreover, these particles are directly measured in tracking detectors, making them ideal for experimental analysis. This motivates the introduction of the track-based NEC in Eq.~\eqref{eq:NEEC_def} and the DIS energy pattern on tracks in Eq.~\eqref{eq:energy pattern} in our work.

However, restricting the measurement to a subset of hadrons introduces non-perturbative effects from hadronization. In this case, the summation over the final-state hadron species in Eq.~(\ref{eq:sum1}) becomes less inclusive, leading to the survival of final-state collinear singularities at the parton level. Consequently, the description of the energy flux depends on parton-to-hadron fragmentation functions (FFs)\footnote{In principle, energy correlators on tracks~\cite{Chen:2020vvp,Li:2021zcf,Jaarsma:2022kdd,Jaarsma:2023ell,Lee:2023npz,Lee:2023tkr} require the track-function formalism~\cite{Chang:2013rca,Chang:2013iba} to describe parton fragmentation into a subset of hadrons. However, for a one-point energy correlator, it is equivalent to use standard FFs.}, introducing additional non-perturbative effects compared to the fully inclusive NEC.

Nevertheless, unlike hadron-level observables that require a full convolution over the functional form of FFs, the energy weighting nature of the NECs effectively reduces this dependence to a non-perturbative number—the first moment of the FFs:
 \begin{align}
    T^\mathbb{S}_{a}\equiv\sum_{h \in \mathbb{S}}
    \int ^{1}_0 z d_{h/a}(z) \dd z~.
        \label{eq:TS}
    \end{align}
    In the following, we will demonstrate that  the track-based NECs (NECs for a subset of hadrons) $f^{t,a}_{1T,\mathbb{S}}(x, \theta)$ has a simple relation to the parton-level NECs
    :
    \begin{align}
    f^{t,a}_{1T,\mathbb{S}}&(x, \theta)=T^\mathbb{S}_{\bar a}f^{t,a}_{1T}(x, \theta)~,
    \label{eq:trackNECsmallx}
    \end{align}  
    where $a=q,\bar q$, and $f^{t,a}_{1T}$ on the right-hand side corresponds to the parton-level inclusive Sivers NEC, derived in Eq.~\eqref{eq:qNEEC_O}. This result implies that the fragmentation effects in the track-based NECs are entirely captured by a single non-perturbative parameter, $T_{a}^{\mathbb{S}}$.

\paragraph*{From Fracture functions to NECs} 
To see how these effects emerge, we first perform the calculation at the hadron level and compute the quark fracture function $u^{q}_h(x, \xi_h, \bs P_{h\perp})$ in the small-$x$ limit. We then apply the energy sum rule in Eq.~\eqref{eq:connection}, which relates fracture functions to NECs. The calculation of the small-$x$ quark fracture functions is similar to the inclusive NEC, except the convolution of FFs.

We find that the small-$x$ quark fracture function for producing a hadron $h$ is expressed in terms of the dipole gluon distribution as follows:
   \begin{align}
    u^{q}_h(&x, \xi_h, \bs P_{h\perp},\bs S_\perp)\notag\\
     =&\frac{1}{\xi_h 8\pi^4}
    \int^1_0\frac{\dd z}{z^2} d_{h/\bar q}(z)\theta(1-x-\xi_h/z)
    \notag \\ 
    &
    \times\int \dd^2\bs k_{g\perp}{\cal H}_q\Big(\frac{\xi_h}{z},\frac{\bs P_{h\perp}}{z},\bs k_{g\perp}\Big) {\cal F}_{x_g}(\bs k_{g\perp},\bs S_\perp)~.
    \end{align} 
Here $d_{h/\bar q}(z)$ represents the antiquark FF into the hadron $h$, and it is convoluted with the hard coefficient function ${\cal H}_q$, given in Eq.~(\ref{eq:Afunction}). 

Applying the energy sum rule in Eq.~\eqref{eq:connection}, the quark NEC for a subset of hadrons $\mathbb{S}$ is given by
\begin{align}
 f^q_{\mathbb{S}}&(x,\theta,\phi, \bs S_\perp) 
 \notag \\ 
=&
\sum_{h\in \mathbb{S} }  \int_0^{1-x} \dd\xi_h\frac{ \xi_h\bs P_{h\perp}^2}{2\theta^2} u^q_h(x,\xi_h, \bs P_{h\perp}, \bs S_\perp)
\Big\vert_{\substack{ \bs P_{h\perp} = \frac{ \xi_h P^+ }{\sqrt{2}}}\bs n_t}~,
\notag \\ 
=&\frac{1}{(2\pi)^4}\sum_{h\in \mathbb{S} }  \int_0^{1-x} \dd\xi_h
\Big(\frac{\xi_h P^+}{\sqrt{2}}\Big)^2
   \int^1_{\frac{\xi_h}{1-z}}\frac{\dd z}{z^2} d_{h/\bar q}(z)
    \notag \\ 
    &
    \times\int \dd^2\bs k_{g\perp}{\cal H}_q\Big(\frac{\xi_h}{z},\frac{ \xi_h P^+ }{z \sqrt{2}}\bs n_t,\bs k_{g\perp}\Big) {\cal F}_{x_g}(\bs k_{g\perp},\bs S_\perp)~.
 \end{align}
Interchanging the integration order between $\xi_h$ and $z$, we obtain: 
 \begin{align}
      f^q_{\mathbb{S}}&(x,\theta,\phi, \bs S_\perp)
      \notag \\ 
      =&\sum_{h \in \mathbb{S}}
\int ^{1}_0 \dd z~z d_{h/\bar q}(z)\frac{1}{(2\pi)^4}
\int^{(1-x)z}_0\frac{\dd \xi_h}{z} 
 \Big (\frac{\xi_h P^+}{z \sqrt{2}} \Big)^2
    \notag \\ 
    &
    \times  \int \dd^2\bs k_{g\perp}{\cal H}_q\Big(\frac{\xi_h}{z},\frac{ \xi_h P^+ }{z \sqrt{2}}\bs n_t,\bs k_{g\perp}\Big) {\cal F}_{x_g}(\bs k_{g\perp},\bs S_\perp)~.
 \end{align}
By changing the integration variable from $\xi_h$ to $\xi_{\bar q}=\xi_h/z$, the $z$-integral factorizes completely. As a result, the quark NEC from a subset of hadrons $\mathbb{S}$ simplifies to the following form:
    \begin{align}
  f^q_{\mathbb{S}}&(x,\theta,\phi, \bs S_\perp)=
  \Big(\sum_{h \in \mathbb{S}}\int ^{1}_0 z d_{h/\bar q}(z) dz\Big)
  f^q(x,\theta,\phi, \bs S_\perp)~,
  \label{eq:fS}
    \end{align}  
    with $f^q(x,\theta,\phi, \bs S_\perp)$ on the right-hand side corresponds to the inclusive NEC derived in Eq.~(\ref{eq:smallNEEC}). A similar relation holds for the antiquark NEC. As a consistency check, the inclusive case is recovered when summing over all hadrons: \begin{align} T^\mathbb{S}_{a}\big \vert_{\mathbb{S}=\text{all hadrons}}=\sum_{h}\int ^{1}_0 \dd z z d_{h/a}(z) =1~.
\end{align}
Here, we have used the momentum sum rule of the FFs.

\paragraph*{SSA for a subset of hadrons}Finally, extracting the $\bs S_\perp$-dependent component in Eq.~\eqref{eq:fS}, we obtain the relation for the Sivers NEC, as given in Eq.~\eqref{eq:trackNECsmallx}. The flavor-singlet Sivers NEC then takes the form: 
    \begin{align}
    \sum_{a=q,\bar q}f^{t,a}_{T, \mathbb{S}}(x, \theta)=\sum_q(T^\mathbb{S}_{\bar q}-T^\mathbb{S}_{q})f^{t,q}_{T}(x, \theta)~.
    \end{align}
      Here, the summation $\sum_q$ run over the quark flavors. 
  This result confirms that to obtain a nonzero flavor-singlet Sivers NEC, we must select a subset of hadrons such as $\mathbb{S}=\{\pi^+\}$ or $\{\pi^-\}$, ensuring $T^\mathbb{S}_{\bar q}-T^\mathbb{S}_{q} \neq 0$.

    Similarly, the $\bs S_\perp$-independent quark NEC can be expressed as
    $f^{a}_{1,\mathbb{S}}(x, \theta)=T_{\bar a}^{\mathbb{S}}f^{a}_{1}(x, \theta)$, and its associated flavor-singlet is
    \begin{align}
    \sum_{a=q,\bar q}f^a_{1, \mathbb{S}}(x, \theta)=\sum_q(T_{\bar q}^{\mathbb{S}}+T_{ q}^{\mathbb{S}})f^q_{1 }(x, \theta)~.
    \label{eq:unfq}
    \end{align}
     With the above results, the SSA for a subset of hadrons $\mathbb{S}$ in Eq.~\eqref{eq:SSANEC} can be expressed as 
     \begin{align}
         A_{UT}^ \mathbb{S}(\theta)=
    {\cal{R}}_{\mathbb{S}} \frac{f^{t,q}_{T}(x_B, \theta)}{f^{q}_1(x_B, \theta)}~,
    \label{eq:O1}
     \end{align}
    where the non-perturbative $\theta$-independent factor is defined as
    \begin{align}
       {\cal{R}}_{\mathbb{S}}= \frac{\sum_q e_q^2(T^\mathbb{S}_{\bar q}-T^\mathbb{S}_{q})}{\sum_q e_q^2(T^\mathbb{S}_{\bar q}+T^\mathbb{S}_{q})}~.
       \label{eq:R_S}
    \end{align}
    For a given subset of hadrons, this factor can be computed in terms of fragmentation functions extracted from fits to experimental data. We are particularly interested in the SSA for positively and negatively charged hadrons. Since the odderon is $C$-odd, we expect the SSA to change sign if we switch the observed energy flow from positively charged hadrons to negatively charged ones. 
    
    To illustrate this, we can use the LO FF set from NNFF1.0~\cite{Bertone:2018ecm,Bertone:2017tyb} for numerical estimates. For charged pions $\mathbb{S}=\{\pi^+\}
    $ or $\{\pi^-\}$, we find ${\cal{R}}_{\{\pi^+\}}\approx -{\cal{R}}_{\{\pi^-\}}=-0.270$ at $\mu=5$ GeV in the four-flavor scheme (with $z_{\text{min}}=0.1$). To avoid  complications from particle identification, 
     we also consider including all positively or negatively charged particles. This yields ${\cal{R}}_{\{h^+\}}\approx -{\cal{R}}_{\{h^-\}}=-0.252$ in NNFF1.0.  Since tracking detectors provide better angular resolution and pointing accuracy than calorimeters, they are well-suited for measuring the energy flow from inclusive charged particles.
    
 While our primary focus is on charged hadrons, the theoretical framework established here applies to any subset of hadrons that satisfies $T^\mathbb{S}_{\bar q}-T^\mathbb{S}_{q}\neq 0$. For  instance, heavy-flavor hadrons could provide complementary insight into the spin-dependent odderon by probing different partonic subprocesses and fragmentation dynamics.

    \section{The charge pattern in DIS}
    \label{sec:charge pattern}
     We now extend our analysis to the charge pattern  and introduce the nucleon charge correlator. Let us first present the motivation for this extension. Although restricting the energy pattern on charged particles effectively facilitate the probe for spin-dependent odderon, it fails to impose a $C$-odd constraint on the final states unless the subset of hadrons $\mathbb{S}$ satisfies $T^\mathbb{S}_{\bar q}=-T^\mathbb{S}_{q}$. It is important to note that the connection between the Sivers NEC and the spin-dependent odderon, established in the previous section, holds in the small-$x$  limit.
     
     In practical applications, particularly when extracting the odderon from EIC data where $x$ is not asymptotically small, the absence of a $C$-odd constraint may lead to contamination from $C$-even gluonic contributions. While the spin-dependent odderon  arises from color-symmetrical tri-gluon correlations $\sim d^{abc} A^{a,+} A^{b,+} A^{c,+}$ in the dilute limit, the $C$-even gluonic contributions can appear through the color-antisymetric tri-gluon correlations $\sim f^{abc} A^{a,+} A^{b,+} A^{c,+}$. For example, it has been explicitly demonstrated in~\cite{Chen:2024bpj} that such $C$-even gluonic correlations can contribute to the Sivers quark NEC  in the region $\Lambda_{QCD}\ll \theta Q \ll Q$ 
     within the twist-3 collinear framework. A consistent matching between the small-$x$ framework and the collinear framework can be only achieved if the $C$-even gluonic correlations vanish in the small-$x$ limit. A detailed analysis of this matching is provided in Appendix~\ref{sec:appendix}.

The $C$-even gluonic contributions can be inherently eliminated by constructing a $C$-odd tag on the final state. We achieve this by incorporating the electric charge of the final-state hadrons into the energy flux measurement. The corresponding observable, the DIS charge pattern, defined in Eq.~\eqref{eq:charge pattern}, provides a cleaner probe of the odderon contribution, serving as a complementary approach to the DIS energy pattern. To describe this observable, we extend the NECs to the nucleon charge correlators
(NCCs). Since charge weighting only affects the final-state hadrons, the analysis follows a similar framework as the energy pattern presented in previous sections. In the following, we present the results directly.

\subsection{Factorization and nucleon charge correlators}
 We first define the charge-weighted energy flow operator as 
 \begin{align} {\cal E}_{\mathbb{Q}}&(\theta,\phi)
 \notag \\ 
 =&\sum_{h} \int \frac{\dd P_h^+ \dd[2]\bs P_{\perp} }{2P_h^+(2\pi)^3}\frac{E_hQ_h}{E_N}\delta(\theta^2-\theta^2_h)\delta(\phi-\phi_h)a^\dagger_h a_h~. \end{align}
By construction, this operator is C-odd, as it assigns opposite signs to positively and negatively charged hadrons due to the presence of the hadronic electric charge $Q_h$~\cite{Lee:2023npz,Lee:2023tkr}:
\begin{align} {\cal E}_{\mathbb{Q}} |h\rangle = E_h Q_h |h\rangle~. \end{align}
With this operator we define the NCC as:
\begin{align}
f^q_{\mathbb{Q}}&(x,\theta,\phi, \bs S_\perp)
\notag \\ 
=&\int  \frac{\dd y^-}{4\pi}e^{-ix P^+y^-} 
\notag \\
&\times
\langle PS_\perp|\bar\psi(y){\cal L}^\dagger(y)\gamma^+{\cal E}_{\mathbb{Q}}(\theta,\phi) {\cal L}(0)\psi(0)|PS_\perp\rangle
\notag \\
=&f^q_{1,\mathbb{Q}}\left(x, \theta\right)+\frac{\epsilon^{\perp}_{\mu \nu} \bs S_\perp^{\mu}\bs {n}_{t}^\nu}{|\bs n_t|} f^{t,q}_{1T,\mathbb{Q}}\left(x,\theta\right)\,.
\label{eq:NEEC_def2}
\end{align}
Here $f^q_{1,\mathbb{Q}}$ and $f^{t,q}_{1T,\mathbb{Q}}$ represent the unpolarized NCC and the Sivers-type NCC, respectively.  Essentially, the NCCs can be understood as the charge-weighted NECs.

In terms of the NCCs, the DIS charge pattern in the TFR follows a similar factorization formula as in Eq.~\eqref{eq:factorization}: \begin{align}
&\frac{\dd \Sigma_{\mathbb{Q}}(\theta, \phi, \bs S_\perp)}{\dd x_B \dd Q^2 }
=\frac{2\alpha_\mathrm{em}^2 }{ Q^4}\sum_{\lambda=L,T}f_\lambda(y)
\notag \\ 
&\quad\quad\times\sum_{a}\int_{x_B}^1 \frac{\dd x}{x} {\cal H}_{a,\lambda}\Big(\frac{x_B}{x},\frac{Q}{\mu}\Big) 
f^a_{\mathbb{Q}}(x,\theta,\phi, \bs S_\perp,\mu)~.
\end{align} 
At LO in $\alpha_s$, the corresponding charge pattern structure functions can be written as
\begin{align}
\Sigma_{UU,T}^{\mathbb{Q}}=&\frac{2\alpha_\mathrm{em}^2 }{ Q^4}f_T(y)\sum_{a=q,\bar q} e_a^2 f^a_{1,\mathbb{Q}}(x_B,\theta)~,
 \\ 
    \Sigma_{U T,T}^{\mathbb{Q}}=&\frac{2\alpha_\mathrm{em}^2 }{ Q^4}f_T(y)\sum_{a=q,\bar q} e_a^2 f^{t,a}_{1T,\mathbb{Q}}(x_B,\theta)~.    
\end{align}
The SSA of the charge pattern is given by the ratio between the flavor singlets of the two NCCs:
\begin{align} &A_{UT}^{\mathbb{Q}}=  \frac{\sum_{a=q,\bar q} e_a^2 f^{t,a}_{1T,\mathbb{Q}}(x_B,\theta) }{\sum_{a=q,\bar q} e_a^2 f^a_{1,\mathbb{Q}}(x_B,\theta)}~. 
\label{eq:SSANCC}
\end{align}

\subsection{Connection to the spin-dependent odderon}
Following the analysis in Sec.~\ref{sec:NECsmallx}, the small-$x$ Sivers NCC $f^{t,a}_{T,\mathbb{Q}}$ can be expressed with the associated parton-level inclusive NEC $f^{t,a}_{T}$: 
\begin{align}
    f^{t,a}_{1T,\mathbb{Q}}(x, \theta)=D_{\bar a}^{\mathbb{Q}}f^{t,a}_{1T}(x, \theta)~,
\end{align}
    where $a=q,\bar q$ and $D_{\bar a}^\mathbb{Q}$ represents the charge weighted moment of the FFs, defined as:
    \begin{align}
    D_{\bar a}^\mathbb{Q}=\sum_{h }Q_h
    \int ^{1}_0 \dd zz d_{h/\bar a}(z)~.
    \end{align}  
 Furthermore, charge conjugation invariance guarantees~\cite{Chen:2020vvp}:
  \begin{align}
      D_{ q}^\mathbb{Q}=-D_{\bar q}^\mathbb{Q}~.
      \label{eq:DQ}
  \end{align}
This effectively establishes a $C$-odd tag on the final states, as the flavor-singlet Sivers NCC is now expressed as the difference between the quark and antiquark NEC:
     \begin{align}
    \sum_{a=q,\bar q}e_a^2f^{t,a}_{1T, \mathbb{Q}}(x, \theta)
    =\sum_q e_q^2 D_{ \bar q}^\mathbb{Q}\big[f^{t,q}_{1T}(x, \theta)-f^{t,\bar q}_{1T}(x, \theta)\big]~,
    \end{align}
  where the summation $\sum_q$ runs over the quark flavors. 

  To understand how the $C$-odd tag works in practice, we note that the above relation remains valid at moderate-$x$, where the twist-3 collinear formalism applies. In this regime, while the quark and antiquark Sivers NCCs receive the same $C$-even contributions, their $C$-odd correlations appear with opposite signs, preserving the charge asymmetry. However, since the flavor-singlet Sivers NCC is determined by the difference between the quark and antiquark NCCs, the $C$-even contributions cancel, leaving only the 
$C$-odd effects.

Now, let us focus on the small-$x$ region, where the spin-dependent odderon dominates. The sign reversal introduced by the final-state charge weighting ($D_{ q}^\mathbb{Q}=-D_{\bar q}^\mathbb{Q}$) exactly cancels with the sign difference from the odderon in the initial state ($f^{t,\bar q}_{1T}(x, \theta)=-f^{t,q}_{1T}(x, \theta)$). Thus, we have 
\begin{align}
    \sum_{a=q,\bar q}e_a^2f^{t,a}_{1T,\mathbb{Q}}(x, \theta)
    =2\sum_q e_q^2 D_{ \bar q}^\mathbb{Q} f^{t,q}_{1T}(x, \theta)~,
    \end{align}
    where we have used the quark-antiquark symmetry relation of the inclusive Sivers NEC given in Eq.~\eqref{eq:fqop}.

  However, the unpolarized NCC $f^q_{1,\mathbb{Q}}$ vanishes, as it is only sensitive to the pomeron contributions, which is C-even. Consequently, the structure function $\Sigma_{UU}^{\mathbb{Q}}$ is not a suitable normalization of the charge-pattern  SSA in Eq.~\eqref{eq:SSANCC}. Instead, we normalize the charge-pattern SSA by the unpolarized energy-pattern structure function  $\Sigma_{UU,T}^{\mathbb{S}}$ from all charged hadrons:
      \begin{align}
        \tilde A_{U T}^\mathbb{Q}(\theta)\equiv&\frac{\Sigma_{UT}^{\mathbb{Q}}}{\Sigma_{UU}^{\mathbb{S}}}
          =\Bigg(\frac{\sum_q e_q^2 D_{ \bar q}^\mathbb{Q}}{\sum_q e_q^2 T_{\bar q}^{\mathbb{S}}}\Bigg) \frac{f^{t,q}_{1T}(x, \theta)}{ f^q_{1}(x, \theta)}~, 
        \label{eq:O2new}
    \end{align}
    where we have used $T_{ q}^{\mathbb{S}}=T_{ \bar q}^{\mathbb{S}}$ for $\mathbb{S}=\text{all}$ charged hadrons. 
For estimation, we use the LO FF set from NNFF1.0~\cite{Bertone:2018ecm,Bertone:2017tyb} which gives $\sum_q e_q^2 D_{ \bar q}^\mathbb{Q}
=-0.041$ and $\sum_q e_q^2 T_{\bar q}^{\mathbb{S}}=0.542$ at $\mu=5$ GeV in the four-flavor scheme.

\section{Predictions for the EIC}
\label{sec:numerics}

We now perform a numerical evaluation of the SSA, defined in Eq.~\eqref{eq:SSAdef}, for the track-based energy pattern, Eq.~\eqref{eq:O1}, and for the charge pattern, Eq.~\eqref{eq:O2new}, and demonstrate that their angular distributions  are sensitive to the odderon. 
We consider energy flux in EIC kinematics with $x_B=3 \times 10^{-3},~ Q^2=25~\mathrm{GeV}^2$ and $\sqrt{s_{ep}}=105~\mathrm{GeV}$ and focus on events where $\theta>0.1$ in the Breit frame. This lower cut ensures separation of the detected forward particles from the target beam remnants and corresponds to a laboratory frame pseudorapidity $\eta_{Lab}<1.8$ at the EIC (here the proton moves into the direction of positive pseudorapidity). These events will be well-covered by the central detector with full azimuthal acceptance~\cite{AbdulKhalek:2021gbh}.

We first compute the baseline of the SSA in Eq.~\eqref{eq:O1}, namely the unpolarized energy pattern $ \Sigma_{UU}$,
corresponding to the $\phi_S$ independent part of the energy pattern~\eqref{eq:SSAdef} including all hadrons,
 or the quark NEC $f_1^q$ defined in Eq.~\eqref{eq:NEEC_def}. As shown in Eq.~(\ref{eq:qNEEC_P}), this NEC only involves the pomeron (dipole-proton) amplitude. For that we use both the GBW model from~\cite{Golec-Biernat:1998zce} and the MVe fit from~\cite{Lappi:2013zma}. In the GBW model the pomeron amplitude in the momentum space reads
\begin{equation}
   F_{x_g}(\bs {k}_{g\perp}^2) = \frac{R^2}{Q_{s,p}^2(x_g)}\exp\big[-\bs k_{g\perp}^2/{Q_{s,p}^2(x_g)}\big].
   \end{equation}
In this model the free parameters are the proton tranverse area $\pi R^2 = 11.51$ mb and the $x_g$-dependent saturation scale 
$Q_{s,p}^2(x_g)= Q_{0}^2 \left(x_0/x_g\right)^\lambda$,
with $Q_{0}^2=1\,\mathrm{GeV}^2$, $x_0=3.04\times 10^{-4}$ and $\lambda=0.288$.
The free parameters are constrained in Ref.~\cite{Golec-Biernat:2017lfv} by fitting the proton structure function data at small-$x$.
Similarly the structure function data is used in Ref.~\cite{Lappi:2013zma} to constrain the initial condition for the small-$x$ running coupling Balitsky-Kovchegov (rcBK) evolution~\cite{Kovchegov:1999yj,Balitsky:1995ub,Balitsky:2006wa} of the dipole-proton scattering amplitude to obtain the ``MVe'' fit. We take $x_g=x_B$ in the calculation.

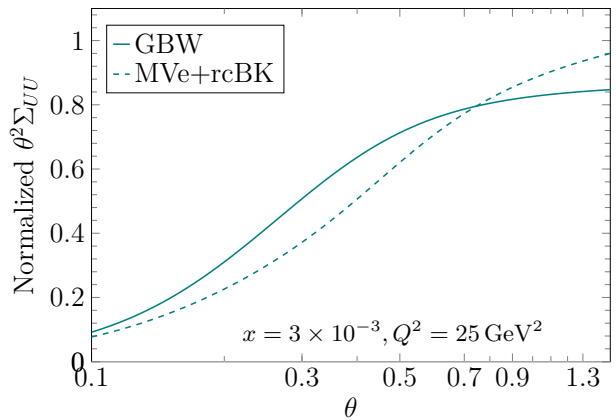
\begin{figure}[tb]
\centering
\includegraphics[width=0.45\textwidth]{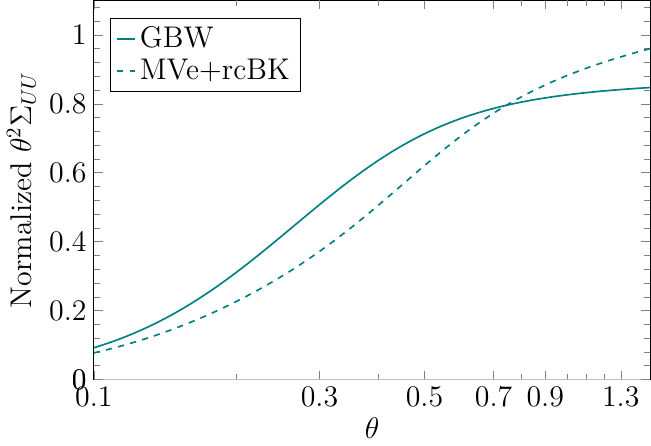}
\begin{tikzpicture}[overlay,remember picture]
\node at (-3,1.2){$x=3 \times 10^{-3}, Q^2=25\,\mathrm{GeV}^2$};
 \end{tikzpicture}
\caption{Unpolarized energy pattern for proton at the EIC
}
\label{fig:unEP}
\end{figure}

In Fig.~\ref{fig:unEP}, the 
$\theta$-distribution of the unpolarized energy pattern $\theta^2 \Sigma_{UU}$ is shown for the two different pomeron amplitudes, normalized by ${\cal N}\equiv \int^{\theta_{\text{max}}}_{\theta_{\text{min}}} \dd \theta   \, \theta ^2 \Sigma_{UU}$ to emphasize the shape rather than the magnitude. This normalization ensures that the energy pattern for any subset of hadrons $\mathbb{S}$ coincides with the inclusive energy pattern when the pomeron input is fixed. This uniformity arises because, as indicated in Eq.~\eqref{eq:unfq}, the non-perturbative track factor affects only the overall normalization, leaving the shape invariant.  The results align well with the inclusive energy patterns reported in Ref.~\cite{Liu:2023aqb}. Both the MVe+rcBK and GBW models used for the pomeron amplitude result in an excellent description of the small-$x$ total DIS cross section data from HERA~\cite{Lappi:2013zma,Golec-Biernat:2017lfv}.
However, numerically significant differences are obtained at the level of the NECs, for example at $\theta=0.3$ the difference is up to $\sim 30\%$.
This difference demonstrates that the future EIC energy pattern measurements can provide complementary constraints to the extraction of the dipole-proton scattering amplitude.

To compute the transversely-polarized energy pattern $\Sigma_{UT}^{\mathbb{S}}$, or the quark Sivers NEC $f_{T,{\mathbb{S}}}^{t}$, the spin-dependent odderon serves as a crucial input. However, unlike the pomeron, the spin-dependent odderon $O_{1 T,x_g}^{\perp}(\bs k_{\perp}^2)$ remains poorly constrained, with limited guidance on its dependence on $x_g$ and $\bs k_\perp$ available in the literature. To explore the sensitivity of the SSA to these inputs, we employ two models in our computations. The first model is the generalized MV model~\cite{Zhou:2013gsa}, where the odderon amplitude is expressed in terms of the pomeron amplitude as follows: 
 \begin{align}
 O_{1 T,x_g}^{\perp}(\bs k_{\perp}^2)=-\frac{c_0 \alpha_s^3\left(\kappa_p^u+\kappa_p^d\right)}{4 R^4}\frac{\partial}{\partial \bs k_{\perp}^2} \frac{\partial}{\partial \bs k_{\perp}^i} \frac{\partial}{\partial \bs  k_{\perp i}} F_{x_g}(\bs k_{\perp}^2)~,
 \label{eq:Omodel1}
 \end{align}
 where $\kappa_p^u=1.673$, $\kappa_p^d=-2.033$ and $c_0=(N_c^2-1)(N_c^2-4)/(4 N_c^3)$.  The only required input in this model is the pomeron amplitude $ F_{x_g}(\bs {k}_{g\perp}^2) $, for which we adopt the GBW and MVe models introduced earlier. The second model is derived from a fit of the gluon Sivers TMD in~\cite{DAlesio:2015fwo}, leveraging the established relation between the gluon Sivers TMD and the odderon~\cite{Boer:2015pni}. In this model, the odderon is parametrized using a Gaussian ansatz 
\begin{align}
 O_{1 T,x_g}^{\perp}( \bs k^2_\perp)=\frac{4 \pi^2 \alpha_sM_p }{ \bs {k}_\perp^3N_c}\mathcal{N}(x_g) g(x_g) h_g(k_{\perp}) \frac{e^{-\bs k_{\perp}^2 /\langle \bs k_{\perp}^2\rangle}}{\pi\langle  \bs k_{\perp}^2\rangle}~,
  \label{eq:Omodel2}
\end{align} 
where $\mathcal{N}\left(x\right)=N x^{\alpha}(1-x)^{\beta}(\alpha+\beta)^{(\alpha+\beta)}/(\alpha^{\alpha} \beta^{\beta}) $, $h(\bs k_{\perp})=\sqrt{2 e} e^{-\bs k_{\perp}^2 / M_g^2}\bs k_{\perp}/M_g $, and $\langle \bs k_{\perp}^2\rangle=0.25~\mathrm{GeV}^2$. Here, $g(x)$ denotes the unpolarized gluon PDF.
Two fits to extract the non-perturbative parameters have been performed in Ref.~\cite{DAlesio:2015fwo}. The first fit is labeled as  ``SIDIS1'', with parameters $N=0.65,~\alpha=2.8,~\beta=2.8$ and $~M_g^2=0.5487~\mathrm{GeV}^2$. The second one, labeled as ``SIDIS2'', is given as $N=0.05,~\alpha=0.8,~\beta=1.4,~M_g^2=0.3396~\mathrm{GeV}^2$. The GRV98-LO set~\cite{Gluck:1998xa} is adopted for $g(x_g)$ in the fits. We evaluate it at the scale $\mu=Q$ and $x_g=x_B$.

\begin{figure}[tb]
\centering
\includegraphics[width=0.45\textwidth]{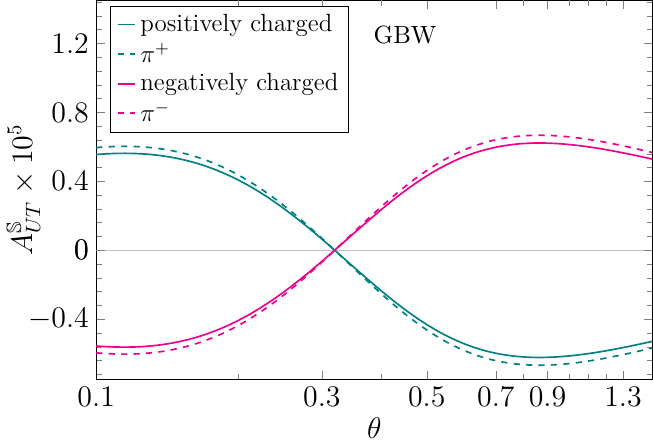}
\begin{tikzpicture}[overlay,remember picture]
\node[scale=0.89] at (-2.0,4.5){$x=3 \times 10^{-3}, Q^2=25\,\mathrm{GeV}^2$};
 \end{tikzpicture}
\caption{The SSA $A_{UT}^{\mathbb{S}}(\theta)$ for the positively and negatively charged particles within the GBW model. Both the transversely-polarized and unpolarized energy pattern are computed with the GBW model as input. 
}
\label{fig:AUTGBW}
\end{figure}
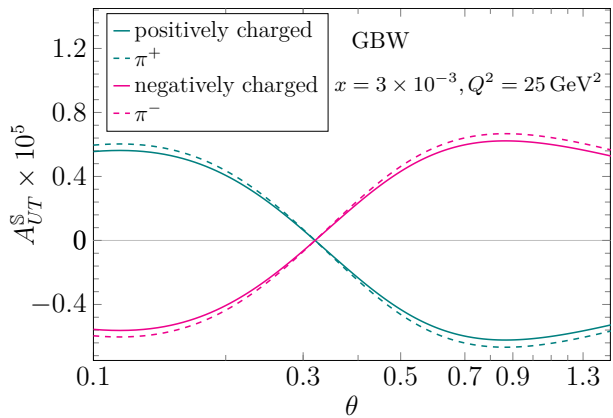

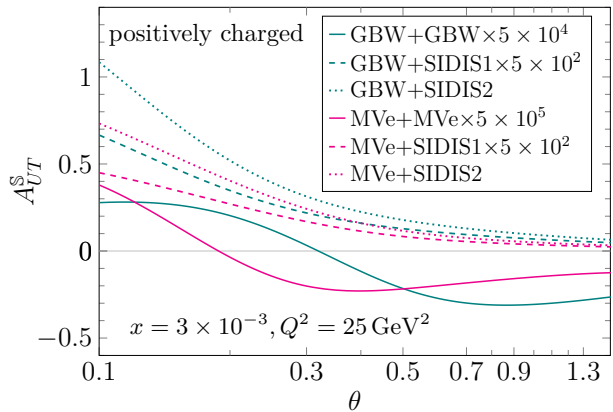
\begin{figure}[tb]
\centering
\includegraphics[width=0.45\textwidth]{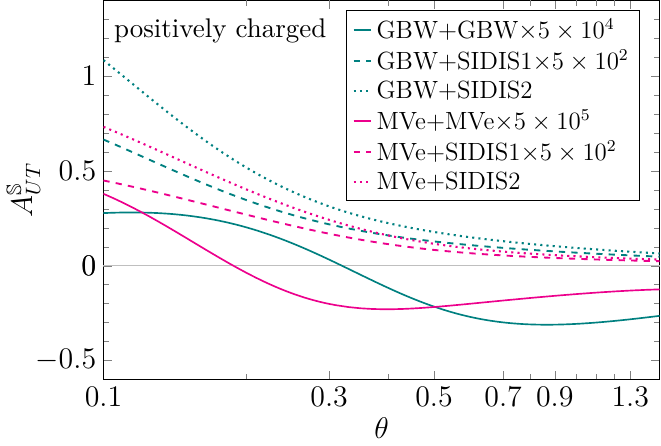}
\begin{tikzpicture}[overlay,remember picture]
\node at (-4.5,1.2){$x=3 \times 10^{-3}, Q^2=25\,\mathrm{GeV}^2$};
 \end{tikzpicture}
\caption{The SSA $A_{UT}^\mathbb{S}(\theta)$ for all positively charged particles. For the pomeron amplitude, the GBW and MVe models are used. The odderon amplitude is obtained in terms of the pomeron amplitude using Eq.~\eqref{eq:Omodel1}, or by using one of the SIDIS fits.
 }
\label{fig:AUT}
\end{figure} 

Fig.~\ref{fig:AUTGBW} illustrates the SSA $A_{UT}^\mathbb{S}$, defined in Eq.~\eqref{eq:SSAdef}, as a function of $\theta$ for positively and negative charged particles, and for $\pi^+$ and $\pi^-$. The results are obtained by calculating the odderon contribution from Eq.~\eqref{eq:Omodel1} using the GBW model for the pomeron.  The associated non-perturbative factor $ {\cal{R}}_{\mathbb{S}}$, given in Eq.~\eqref{eq:R_S}, is evaluated using the LO FF set from NNFF1.0~\cite{Bertone:2017tyb} at $\mu=Q$. A key feature in the figure is the sign reversal of the SSA between positively and negatively charged particles, reflecting the C-odd nature of the spin-dependent odderon. This sign-changing behavior is consistent with the non-perturbative factor ${\cal R}_{\mathbb{S}}$ derived from the fragmentation functions, and serves as a direct experimental signature of the odderon.

\begin{figure}[tb]
\centering
\includegraphics[width=0.45\textwidth]{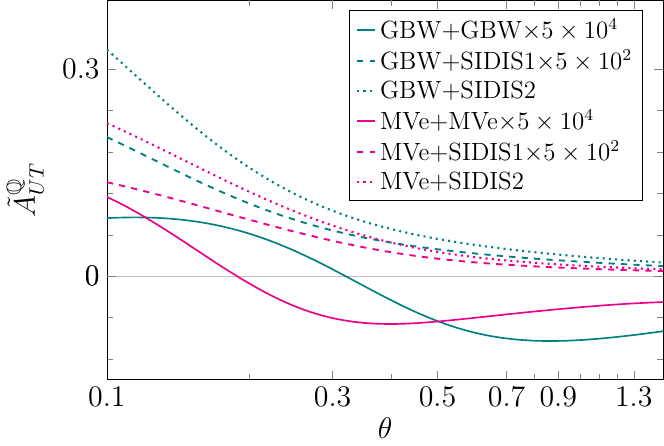}
\begin{tikzpicture}[overlay,remember picture]
\node at (-4.5,1.2){$x=3 \times 10^{-3}, Q^2=25\,\mathrm{GeV}^2$};
 \end{tikzpicture}
\caption{The SSA $\tilde A_{UT}^\mathbb{Q}(\theta)$ with different models of pomeron and odderon amplitudes as inputs. 
 }
\label{fig:AQUT}
\end{figure}
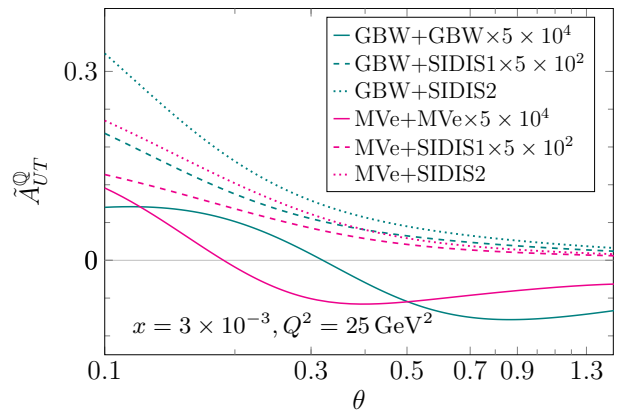 

In Fig.~\ref{fig:AUT}, we present the SSA $A_{UT}^\mathbb{S}$ for positively charged hadrons, computed using different models for the odderon and pomeron inputs. 
The odderon-independent reference (denominator in Eq.~\eqref{eq:SSAdef}) is calculated using both the GBW and MVe models for the pomeron amplitude. With both references, the odderon amplitude is calculated using three different setups. 
First, we compute odderon from the corresponding pomeron amplitude by applying Eq.~\eqref{eq:Omodel1}. These results are referred to as ``GBW+GBW'' and ``MVe+MVe'' in Fig.~\ref{fig:AUT}. For comparison, we also use the two SIDIS fits of the form~\eqref{eq:Omodel2} discussed above.

The results reveal a pronounced sensitivity to the choice of odderon model. While all predictions in Fig.~\ref{fig:AUT} indicate positive SSAs at small $\theta$ ($\theta<0.2$), the SSAs computed with the odderon amplitude obtained by applying Eq.~\eqref{eq:Omodel1} (solid lines), exhibit a characteristic node signifying a sign change occurring between $\theta=0.2$ and $\theta=0.3$. In contrast, the SSA based on the SIDIS fits for the odderon, Eq.~\eqref{eq:Omodel2}, remain positive across all $\theta$ and diminishes rapidly at larger angles. The shapes of these distributions are also distinct in the small-$\theta$ region. 
Furthermore, the overall normalization predicted for the asymmetry is strongly model dependent.
As such,  $A_{UT}^{\mathbb{S}}$ can be identified as a powerful probe of the spin-dependent odderon dynamics at small $x$.

Fig.~\ref{fig:AQUT} presents the $\theta$-distributions of the charge-pattern SSA $\tilde A^{\mathbb Q}_{UT}$. The results exhibit a similar pattern to the energy-pattern SSA $A^{\mathbb S}_{UT}$ shown in Fig.~\ref{fig:AUT}, demonstrating that $\tilde  A^{\mathbb Q}_{UT}$ is highly sensitive to the spin-dependent odderon. Numerically, we find that $\tilde A^{\mathbb Q}_{UT} \approx e A^{\mathbb S}_{UT}$ for ${\mathbb S }=\{h^+\}$, assuming odderon dominance and considering the primary charged hadron species ${\mathbb S}=\{\pi^+, K^+,p\}$ in predicting $A^{\mathbb S}_{UT}$.
However, as we discussed in Sec.~\ref{sec:charge pattern}, the charge-pattern SSA $\tilde  A^{\mathbb Q}_{UT}$ has a crucial advantage. By construction, it is sensitive exclusively to $C$-odd contributions, making it a cleaner probe for the spin-dependent odderon. This is particularly important when approaching the onset of small-$x$ dynamics, where contamination from $C$-even gluonic exchanges might obscure the signal in the energy-pattern observables $A^{\mathbb S}_{UT}$. As a result, the charge-pattern SSA provides a more robust and unambiguous experimental handle on the odderon. The upcoming EIC offers an ideal platform to test these predictions and deepen our understanding of the odderon’s role in the proton’s spin structure.

\section{Conclusions}

\label{sec:conclu}
In summary, we have explored the SSA in the energy pattern of hadrons produced in the TFR of DIS. Our focus has been on investigating the connection between the T-odd quark NEC and the spin-dependent odderon at small-$x$. By introducing track-based NECs, we demonstrated how restricting the measurement to charged hadrons leads to a nonzero SSA. To formalize this connection, we derived the track-based NECs from fracture functions using the energy sum rule. Unlike hadron-level observables, where fragmentation effects enter through a convolution, the energy-weighting nature of this observable ensures that non-perturbative fragmentation effects appear only as an overall factor. By selecting an appropriate subset of hadrons, this non-perturbative factor facilitates the probe of odderon contributions.

A key prediction of our framework is the sign reversal of the SSA when switching between positively and negatively charged hadrons, which provides a robust test of the odderon’s C-odd nature. Furthermore, we extend our analysis to charge-weighted NECs, introducing the charge-pattern SSA as a cleaner probe of the spin-dependent odderon. This approach inherently removes contamination from C-even gluonic contributions by constructing a C-odd tag in the final state, thereby enhancing sensitivity to odderon dynamics. Our numerical predictions, based on various odderon models, illustrate the strong sensitivity of the SSAs to the choice of input. This provides a promising avenue for constraining the odderon at the EIC.

 Our work represents a first step toward understanding the spin-dependent odderon through SSAs in TFR. In this study, we have focused on quark NECs within the eikonal approximation. In future work, we aim to extend this framework by exploring the role of gluonic NECs and fracture functions in probing the spin-dependent odderon. It would also be of great interest to investigate sub-eikonal contributions to both quark and gluonic NECs, building on techniques developed in the context of TMDs (see~\cite{Santiago:2024iem} and references therein).

\begin{acknowledgments}
We thank Yoshikazu Hagiwara for collaboration at the early stage of this project, and Jian-Ping Ma, Feng Yuan for inspiring discussions and comments. We also thank Jian Zhou, Sanjin Benić, Felix Hekhorn, Yu Shi and Du-Xin Zheng for helpful conversations. This work is supported by the Research Council of Finland, the Centre of Excellence in Quark Matter and projects 338263 and 359902, and by the European Research Council (ERC, grant agreements No. ERC-2023-101123801 GlueSatLight and No. ERC-2018-ADG-835105 YoctoLHC). The content of this article does not reflect the official opinion of the European Union and responsibility for the information and views expressed therein lies entirely with the authors.
\end{acknowledgments}

\appendix

\section{Matching on the collinear twist-3 formalism: the dilute limit}\label{sec:appendix}
In this appendix, we investigate the matching of the quark Sivers NEC on the collinear twist-3 formalism: the dilute limit.
We have demonstrated that the quark Sivers NEEC originates from the C-odd odderon within the CGC formalism in the small-$x$ region, where $Q_s(x)\gg \Lambda_{\text{QCD}}$. Meanwhile, it has been shown in~\cite{Chen:2024bpj} that the quark Sivers NEEC can also be factorized in terms of collinear twist-3 distributions for the large-$\theta$ regime, $\theta Q \gg \Lambda_{\text{QCD}}$. Unlike the CGC formalism, the collinear approach incorporates both C-odd and C-even correlations, providing a framework for describing SSA over a broader range of $x$.  However, both formalisms are expected to converge in the intermediate region $\theta Q \gg Q_s(x) \gg \Lambda_{\text{QCD}}$ at small-$x$, where the dilute approximation in the CGC framework or the double-logarithm approximation in the collinear framework becomes valid. This convergence provides a valuable opportunity to study the transition of SSA from large-$x$ to small-$x$. The analysis is similar to that for the quark Sivers TMD~\cite{Dong:2018wsp}.

\paragraph*{CGC Formalism in the dilute limit} Starting from the CGC result in Eq.~(\ref{eq:smallNEEC}), we derive the dilute limit by performing a Taylor expansion of the hard factor ${\cal T}_q$, applying the power counting $\bs k_{g\perp} \sim Q_s \ll \bs k_{\perp}$. The leading non-vanishing contribution to the quark Sivers NEEC arises from the third derivative term:
  \begin{align}
& \frac{\epsilon^{\perp}_{\mu \nu} \bs S_\perp^{\mu}\bs {n}_{t}^\nu}{|\bs n_t|}f^{t,q}_{1T}(x, \theta)
\notag \\ 
=&\frac{N_c}{\theta^2 (2\pi)^4 }
\int_0^{1-x}\dd \xi~\bs k_\perp^2
\frac{1}{6} \frac{\partial^3 {\cal H}_q\left(\bs k_\perp, \bs k_{g\perp}\right)}{\partial \bs k_{g\perp}^\alpha \partial \bs k_{g\perp}^\beta \partial \bs k_{g\perp}^\gamma} \Big \vert_{\bs k_{g\perp}=0}
\notag \\ 
&\times  \int \dd[2]\bs k_{g\perp} \bs k_{g\perp}^\alpha \bs k_{g\perp}^\beta \bs k_{g\perp}^\gamma
\frac{\epsilon_{\perp}^{i j} \bs S_{\perp i} \bs k_{g\perp j}}{M}  O_{1 T,x_g}^{\perp}( \bs  k_{g\perp}^2)~.
\end{align}
The $\bs k_\perp$-moment of the spin-dependent odderon is related to the collinear C-odd tri-gluon correlation functions $O^{\alpha \beta \gamma}\left(x_g\right)$ \cite{Zhou:2013gsa} :
\begin{align}
&\frac{i g^2 \pi^2}{2 N_c} O^{\alpha \beta \gamma}\left(x_g\right)= 
\notag \\ 
&\quad\quad\int \dd[2]\bs k_{g\perp} \bs k_{g\perp}^\alpha \bs k_{g\perp}^\beta \bs k_{g\perp}^\gamma
\frac{\epsilon_{\perp}^{i j} \bs S_{\perp i} \bs k_{g\perp j}}{M}  O_{1 T,x_g}^{\perp}( \bs  k_{g\perp}^2)~
\label{eq:tri1}
\end{align}
where   
\begin{align}
O^{\alpha\beta\gamma}(x_g) =& -2 iO(x_g) \bigr [  g^{\alpha\beta}_\perp \tilde S_\perp^\gamma + 
 g^{\beta\gamma}_\perp \tilde S_\perp^\alpha +g^{\gamma\alpha}_\perp \tilde S_\perp^\beta \bigr ] ~.
 \label{eq:tri2}
\end{align}
The C-odd trigluon correlation functions  are defined as 
\begin{align}
 &O^{\alpha\beta\gamma}(x_1,x_2) = \frac{i^3 g_s}{P^+}  \int \frac{\dd\lambda_1}{2\pi }\frac{\dd\lambda_2}{2\pi} 
   e^{i\lambda_1 x_1 P^+ + i\lambda_2 (x_2-x_1)P^+} 
   \notag  \\ 
   &~~~~\langle PS  \vert d^{abc} F^{a,+\alpha} 
   (\lambda_1 n)  F^{c,+\gamma}(\lambda_2 n) F^{b,+\beta} (0) \vert PS  \rangle ~.
 \label{eq:trigluonf}
\end{align}
Using the leading logarithmic approximation, with $x_g \equiv \operatorname{Max}\left\{x_1, x_2\right\}$ and performing the integration over $z$,  we find that 
 \begin{align}
f^{t,q}_{T}(x, \theta)
=&-\frac{\alpha_s}{ x \pi \theta^3  \sqrt{2} P^+}
 \frac{4}{3}O\left(x_g\right)~,
 \label{eq:matching1}
\end{align}
which is valid in the dilute limit.

\paragraph*{Collinear twist-3 formalism}
In the collinear formalism, the quark Sivers NEC can be expressed in terms of twist-3 quark-gluon and tri-gluon correlations~\cite{Chen:2024bpj}. In the small-$x$ region, the tri-gluon correlations dominate::
\begin{align}
 & f_{1T}^{t,q}(x,\theta) = \frac{\alpha_s }{x\pi \theta^3 \sqrt{2}P^+} \int_x^1  \dd z 
 \big[z^2+(1-z)^2]
 \notag \\ 
 &~~~\Big[N(x_g,x_g)-  N(x_g,0 ) - O(x_g,0)-O(x_g,x_g)\Big]~.
\end{align}
where $x_g=x/z$. Here, $O(x_1, x_2)$ is the C-odd tri-gluon correlation function as defined in Eq.~(\ref{eq:trigluonf}), while $N(x_1, x_2)$ represents the C-even counterpart, obtained by replacing the color tensor $d^{abc}$ with $if^{abc}$ in Eq.~(\ref{eq:trigluonf}). Although both correlations contribute, the C-odd $O(x_1, x_2)$ dominates in the small-$x$ regime because it evolves with an enhanced $1/x_g$ dependence, compared to the C-even $N(x_1, x_2)$~\cite{Schafer:2013opa}. This justifies neglecting the C-even contributions at small-$x$. Thus, we approximate:
\begin{align}
f_{1T}^{t,q}(x,\theta) \approx & - \frac{\alpha_s }{x\pi \theta^3 \sqrt{2}P^+} \int_x^1  \dd z [z^2+(1-z)^2] 2 O(x_g)~.
\end{align}
where $O(x_g,0) \approx O(x_g,x_g)\approx O(x_g)$. Integrating over $z$ confirms that the result matches Eq.~(\ref{eq:matching1}), demonstrating consistency between the CGC and collinear frameworks in the dilute limit.
 
\bibliographystyle{JHEP-2modlong}
\bibliography{ref.bib}

\providecommand{\href}[2]{#2}\begingroup\raggedright\begin{thebibliography}{100}

\bibitem{Lukaszuk:1973nt}
L.~Lukaszuk and B.~Nicolescu, {\it {A Possible interpretation of p p rising
  total cross-sections}},  \href{http://dx.doi.org/10.1007/BF02824484}{{\em
  Lett. Nuovo Cim.} {\bf 8} (1973) 405}.

\bibitem{Donnachie:1985iz}
A.~Donnachie and P.~V. Landshoff, {\it {Dynamics of Elastic Scattering}},
  \href{http://dx.doi.org/10.1016/0550-3213(86)90137-9}{{\em Nucl. Phys. B}
  {\bf 267} (1986) 690}.

\bibitem{Ewerz:2003xi}
C.~Ewerz, {\it {The Odderon in quantum chromodynamics}},
  \href{http://arXiv.org/abs/hep-ph/0306137}{{\tt arXiv:hep-ph/0306137}}.

\bibitem{Kovchegov:2003dm}
Y.~V. Kovchegov, L.~Szymanowski and S.~Wallon, {\it {Perturbative odderon in
  the dipole model}},
  \href{http://dx.doi.org/10.1016/j.physletb.2004.02.036}{{\em Phys. Lett. B}
  {\bf 586} (2004) 267} [\href{http://arXiv.org/abs/hep-ph/0309281}{{\tt
  arXiv:hep-ph/0309281}}].

\bibitem{Hatta:2005as}
Y.~Hatta, E.~Iancu, K.~Itakura and L.~McLerran, {\it {Odderon in the color
  glass condensate}},
  \href{http://dx.doi.org/10.1016/j.nuclphysa.2005.05.163}{{\em Nucl. Phys. A}
  {\bf 760} (2005) 172} [\href{http://arXiv.org/abs/hep-ph/0501171}{{\tt
  arXiv:hep-ph/0501171}}].

\bibitem{Jeon:2005cf}
S.~Jeon and R.~Venugopalan, {\it {A Classical Odderon in QCD at high
  energies}},  \href{http://dx.doi.org/10.1103/PhysRevD.71.125003}{{\em Phys.
  Rev. D} {\bf 71} (2005) 125003}
  [\href{http://arXiv.org/abs/hep-ph/0503219}{{\tt arXiv:hep-ph/0503219}}].

\bibitem{D0:2020tig}
{\bf D0, TOTEM} collaboration, V.~M. Abazov {\em et.~al.}, {\it {Odderon
  Exchange from Elastic Scattering Differences between $pp$ and $p \bar{p}$
  Data at 1.96~TeV and from pp Forward Scattering Measurements}},
  \href{http://dx.doi.org/10.1103/PhysRevLett.127.062003}{{\em Phys. Rev.
  Lett.} {\bf 127} (2021)~no.~6 062003}
  [\href{http://arXiv.org/abs/2012.03981}{{\tt arXiv:2012.03981 [hep-ex]}}].

\bibitem{D0:2012erd}
{\bf D0} collaboration, V.~M. Abazov {\em et.~al.}, {\it {Measurement of the
  differential cross section $d\sigma/dt$ in elastic $p\bar{p}$ scattering at
  $\sqrt{s}=1.96$ TeV}},
  \href{http://dx.doi.org/10.1103/PhysRevD.86.012009}{{\em Phys. Rev. D} {\bf
  86} (2012) 012009} [\href{http://arXiv.org/abs/1206.0687}{{\tt
  arXiv:1206.0687 [hep-ex]}}].

\bibitem{TOTEM:2017sdy}
{\bf TOTEM} collaboration, G.~Antchev {\em et.~al.}, {\it {First determination
  of the ${\rho }$ parameter at ${\sqrt{s} = 13}$ TeV: probing the existence of
  a colourless C-odd three-gluon compound state}},
  \href{http://dx.doi.org/10.1140/epjc/s10052-019-7223-4}{{\em Eur. Phys. J. C}
  {\bf 79} (2019)~no.~9 785} [\href{http://arXiv.org/abs/1812.04732}{{\tt
  arXiv:1812.04732 [hep-ex]}}].

\bibitem{TOTEM:2018psk}
{\bf TOTEM} collaboration, G.~Antchev {\em et.~al.}, {\it {Elastic differential
  cross-section ${\mathrm{d}}\sigma /{\mathrm{d}}t$ at $\sqrt{s}=2.76\hbox {
  TeV}$ and implications on the existence of a colourless C-odd three-gluon
  compound state}},
  \href{http://dx.doi.org/10.1140/epjc/s10052-020-7654-y}{{\em Eur. Phys. J. C}
  {\bf 80} (2020)~no.~2 91} [\href{http://arXiv.org/abs/1812.08610}{{\tt
  arXiv:1812.08610 [hep-ex]}}].

\bibitem{Martynov:2017zjz}
E.~Martynov and B.~Nicolescu, {\it {Did TOTEM experiment discover the
  Odderon?}},  \href{http://dx.doi.org/10.1016/j.physletb.2018.01.054}{{\em
  Phys. Lett. B} {\bf 778} (2018) 414}
  [\href{http://arXiv.org/abs/1711.03288}{{\tt arXiv:1711.03288 [hep-ph]}}].

\bibitem{Lappi:2016gqe}
T.~Lappi, A.~Ramnath, K.~Rummukainen and H.~Weigert, {\it {JIMWLK evolution of
  the odderon}},  \href{http://dx.doi.org/10.1103/PhysRevD.94.054014}{{\em
  Phys. Rev. D} {\bf 94} (2016)~no.~5 054014}
  [\href{http://arXiv.org/abs/1606.00551}{{\tt arXiv:1606.00551 [hep-ph]}}].

\bibitem{Boer:2018vdi}
D.~Boer, T.~Van~Daal, P.~J. Mulders and E.~Petreska, {\it {Directed flow from
  C-odd gluon correlations at small $x$}},
  \href{http://dx.doi.org/10.1007/JHEP07(2018)140}{{\em JHEP} {\bf 07} (2018)
  140} [\href{http://arXiv.org/abs/1805.05219}{{\tt arXiv:1805.05219
  [hep-ph]}}].

\bibitem{Hagiwara:2020mqb}
Y.~Hagiwara, Y.~Hatta, R.~Pasechnik and J.~Zhou, {\it {Spin-dependent Pomeron
  and Odderon in elastic proton-proton scattering}},
  \href{http://dx.doi.org/10.1140/epjc/s10052-020-8007-6}{{\em Eur. Phys. J. C}
  {\bf 80} (2020)~no.~5 427} [\href{http://arXiv.org/abs/2003.03680}{{\tt
  arXiv:2003.03680 [hep-ph]}}].

\bibitem{Dumitru:2021tqp}
A.~Dumitru, H.~M\"antysaari and R.~Paatelainen, {\it {Cubic color charge
  correlator in a proton made of three quarks and a gluon}},
  \href{http://dx.doi.org/10.1103/PhysRevD.105.036007}{{\em Phys. Rev. D} {\bf
  105} (2022)~no.~3 036007} [\href{http://arXiv.org/abs/2106.12623}{{\tt
  arXiv:2106.12623 [hep-ph]}}].
\newblock [Erratum: Phys.Rev.D 109, 119901 (2024)].

\bibitem{Dumitru:2022ooz}
A.~Dumitru, H.~M\"antysaari and R.~Paatelainen, {\it {Stronger C-odd color
  charge correlations in the proton at higher energy}},
  \href{http://dx.doi.org/10.1103/PhysRevD.107.L011501}{{\em Phys. Rev. D} {\bf
  107} (2023)~no.~1 L011501} [\href{http://arXiv.org/abs/2210.05390}{{\tt
  arXiv:2210.05390 [hep-ph]}}].

\bibitem{Benic:2023ybl}
S.~Beni\'c, D.~Horvati\'c, A.~Kaushik and E.~A. Vivoda, {\it {Exclusive
  \ensuremath{\eta}c production from small-x evolved Odderon at an electron-ion
  collider}},  \href{http://dx.doi.org/10.1103/PhysRevD.108.074005}{{\em Phys.
  Rev. D} {\bf 108} (2023)~no.~7 074005}
  [\href{http://arXiv.org/abs/2306.10626}{{\tt arXiv:2306.10626 [hep-ph]}}].

\bibitem{Benic:2024pqe}
S.~Beni\'c, A.~Dumitru, A.~Kaushik, L.~Motyka and T.~Stebel, {\it
  {Photon-odderon interference in exclusive \ensuremath{\chi}c charmonium
  production at the Electron-Ion Collider}},
  \href{http://dx.doi.org/10.1103/PhysRevD.110.014025}{{\em Phys. Rev. D} {\bf
  110} (2024)~no.~1 014025} [\href{http://arXiv.org/abs/2402.19134}{{\tt
  arXiv:2402.19134 [hep-ph]}}].

\bibitem{Benic:2025okp}
S.~Beni\'c and E.~A. Vivoda, {\it {Single spin asymmetry in forward $pA$
  collisions from Pomeron-Odderon interference}},
  \href{http://arXiv.org/abs/2501.12847}{{\tt arXiv:2501.12847 [hep-ph]}}.

\bibitem{Dumitru:2018vpr}
A.~Dumitru, G.~A. Miller and R.~Venugopalan, {\it {Extracting many-body color
  charge correlators in the proton from exclusive DIS at large Bjorken x}},
  \href{http://dx.doi.org/10.1103/PhysRevD.98.094004}{{\em Phys. Rev. D} {\bf
  98} (2018)~no.~9 094004} [\href{http://arXiv.org/abs/1808.02501}{{\tt
  arXiv:1808.02501 [hep-ph]}}].

\bibitem{Zhou:2013gsa}
J.~Zhou, {\it {Transverse single spin asymmetries at small x and the anomalous
  magnetic moment}},  \href{http://dx.doi.org/10.1103/PhysRevD.89.074050}{{\em
  Phys. Rev. D} {\bf 89} (2014)~no.~7 074050}
  [\href{http://arXiv.org/abs/1308.5912}{{\tt arXiv:1308.5912 [hep-ph]}}].

\bibitem{Boer:2015pni}
D.~Boer, M.~G. Echevarria, P.~Mulders and J.~Zhou, {\it {Single spin
  asymmetries from a single Wilson loop}},
  \href{http://dx.doi.org/10.1103/PhysRevLett.116.122001}{{\em Phys. Rev.
  Lett.} {\bf 116} (2016)~no.~12 122001}
  [\href{http://arXiv.org/abs/1511.03485}{{\tt arXiv:1511.03485 [hep-ph]}}].

\bibitem{Szymanowski:2016mbq}
L.~Szymanowski and J.~Zhou, {\it {The spin dependent odderon in the diquark
  model}},  \href{http://dx.doi.org/10.1016/j.physletb.2016.06.055}{{\em Phys.
  Lett. B} {\bf 760} (2016) 249} [\href{http://arXiv.org/abs/1604.03207}{{\tt
  arXiv:1604.03207 [hep-ph]}}].

\bibitem{Hatta:2016wjz}
Y.~Hatta, B.-W. Xiao, S.~Yoshida and F.~Yuan, {\it {Single Spin Asymmetry in
  Forward $pA$ Collisions}},
  \href{http://dx.doi.org/10.1103/PhysRevD.94.054013}{{\em Phys. Rev. D} {\bf
  94} (2016)~no.~5 054013} [\href{http://arXiv.org/abs/1606.08640}{{\tt
  arXiv:1606.08640 [hep-ph]}}].

\bibitem{Dong:2018wsp}
H.~Dong, D.-X. Zheng and J.~Zhou, {\it {Sea quark Sivers distribution}},
  \href{http://dx.doi.org/10.1016/j.physletb.2018.11.010}{{\em Phys. Lett. B}
  {\bf 788} (2019) 401} [\href{http://arXiv.org/abs/1805.09479}{{\tt
  arXiv:1805.09479 [hep-ph]}}].

\bibitem{Yao:2018vcg}
X.~Yao, Y.~Hagiwara and Y.~Hatta, {\it {Computing the gluon Sivers function at
  small-$x$}},  \href{http://dx.doi.org/10.1016/j.physletb.2019.01.029}{{\em
  Phys. Lett. B} {\bf 790} (2019) 361}
  [\href{http://arXiv.org/abs/1812.03959}{{\tt arXiv:1812.03959 [hep-ph]}}].

\bibitem{Boussarie:2019vmk}
R.~Boussarie, Y.~Hatta, L.~Szymanowski and S.~Wallon, {\it {Probing the Gluon
  Sivers Function with an Unpolarized Target: GTMD Distributions and the
  Odderons}},  \href{http://dx.doi.org/10.1103/PhysRevLett.124.172501}{{\em
  Phys. Rev. Lett.} {\bf 124} (2020)~no.~17 172501}
  [\href{http://arXiv.org/abs/1912.08182}{{\tt arXiv:1912.08182 [hep-ph]}}].

\bibitem{Kovchegov:2020kxg}
Y.~V. Kovchegov and M.~G. Santiago, {\it {Lensing mechanism meets small- $x$
  physics: Single transverse spin asymmetry in $p^{\uparrow}+p$ and
  $p^{\uparrow}+A$ collisions}},
  \href{http://dx.doi.org/10.1103/PhysRevD.102.014022}{{\em Phys. Rev. D} {\bf
  102} (2020)~no.~1 014022} [\href{http://arXiv.org/abs/2003.12650}{{\tt
  arXiv:2003.12650 [hep-ph]}}].

\bibitem{Kovchegov:2021iyc}
Y.~V. Kovchegov and M.~G. Santiago, {\it {Quark sivers function at small $x$:
  spin-dependent odderon and the sub-eikonal evolution}},
  \href{http://dx.doi.org/10.1007/JHEP11(2021)200}{{\em JHEP} {\bf 11} (2021)
  200} [\href{http://arXiv.org/abs/2108.03667}{{\tt arXiv:2108.03667
  [hep-ph]}}].
\newblock [Erratum: JHEP 09, 186 (2022)].

\bibitem{Kovchegov:2022kyy}
Y.~V. Kovchegov and M.~G. Santiago, {\it {T-odd leading-twist quark TMDs at
  small x}},  \href{http://dx.doi.org/10.1007/JHEP11(2022)098}{{\em JHEP} {\bf
  11} (2022) 098} [\href{http://arXiv.org/abs/2209.03538}{{\tt arXiv:2209.03538
  [hep-ph]}}].

\bibitem{Boer:2022njw}
D.~Boer, Y.~Hagiwara, J.~Zhou and Y.-j. Zhou, {\it {Scale evolution of T-odd
  gluon TMDs at small x}},
  \href{http://dx.doi.org/10.1103/PhysRevD.105.096017}{{\em Phys. Rev. D} {\bf
  105} (2022)~no.~9 096017} [\href{http://arXiv.org/abs/2203.00267}{{\tt
  arXiv:2203.00267 [hep-ph]}}].

\bibitem{Benic:2024fbf}
S.~Benic, A.~Dumitru, L.~Motyka and T.~Stebel, {\it {Gluon Sivers function from
  forward exclusive \ensuremath{\chi}c1 photoproduction on unpolarized
  protons}},  \href{http://dx.doi.org/10.1103/PhysRevD.111.054008}{{\em Phys.
  Rev. D} {\bf 111} (2025)~no.~5 054008}
  [\href{http://arXiv.org/abs/2407.04968}{{\tt arXiv:2407.04968 [hep-ph]}}].

\bibitem{Zhu:2024iwa}
S.~Zhu, D.~Zheng, L.~Xia and Y.~Zhang, {\it {Charm Sivers function at EicC}},
  \href{http://arXiv.org/abs/2409.00653}{{\tt arXiv:2409.00653 [hep-ph]}}.

\bibitem{Kovchegov:2012ga}
Y.~V. Kovchegov and M.~D. Sievert, {\it {A New Mechanism for Generating a
  Single Transverse Spin Asymmetry}},
  \href{http://dx.doi.org/10.1103/PhysRevD.86.034028}{{\em Phys. Rev. D} {\bf
  86} (2012) 034028} [\href{http://arXiv.org/abs/1201.5890}{{\tt
  arXiv:1201.5890 [hep-ph]}}].
\newblock [Erratum: Phys.Rev.D 86, 079906 (2012)].

\bibitem{Sivers:1989cc}
D.~W. Sivers, {\it {Single Spin Production Asymmetries from the Hard Scattering
  of Point-Like Constituents}},
  \href{http://dx.doi.org/10.1103/PhysRevD.41.83}{{\em Phys. Rev. D} {\bf 41}
  (1990) 83}.

\bibitem{Accardi:2012qut}
A.~Accardi {\em et.~al.}, {\it {Electron Ion Collider: The Next QCD Frontier}:
  {Understanding the glue that binds us all}},
  \href{http://dx.doi.org/10.1140/epja/i2016-16268-9}{{\em Eur. Phys. J. A}
  {\bf 52} (2016)~no.~9 268} [\href{http://arXiv.org/abs/1212.1701}{{\tt
  arXiv:1212.1701 [nucl-ex]}}].

\bibitem{Aschenauer:2017jsk}
E.~C. Aschenauer, S.~Fazio, J.~H. Lee, H.~Mantysaari, B.~S. Page, B.~Schenke,
  T.~Ullrich, R.~Venugopalan and P.~Zurita, {\it {The electron\textendash{}ion
  collider: assessing the energy dependence of key measurements}},
  \href{http://dx.doi.org/10.1088/1361-6633/aaf216}{{\em Rept. Prog. Phys.}
  {\bf 82} (2019)~no.~2 024301} [\href{http://arXiv.org/abs/1708.01527}{{\tt
  arXiv:1708.01527 [nucl-ex]}}].

\bibitem{AbdulKhalek:2021gbh}
R.~Abdul~Khalek {\em et.~al.}, {\it {Science Requirements and Detector Concepts
  for the Electron-Ion Collider}: {EIC Yellow Report}},
  \href{http://dx.doi.org/10.1016/j.nuclphysa.2022.122447}{{\em Nucl. Phys. A}
  {\bf 1026} (2022) 122447} [\href{http://arXiv.org/abs/2103.05419}{{\tt
  arXiv:2103.05419 [physics.ins-det]}}].

\bibitem{Trentadue:1993ka}
L.~Trentadue and G.~Veneziano, {\it {Fracture functions: An Improved
  description of inclusive hard processes in QCD}},
  \href{http://dx.doi.org/10.1016/0370-2693(94)90292-5}{{\em Phys. Lett. B}
  {\bf 323} (1994) 201}.

\bibitem{Grazzini:1997ih}
M.~Grazzini, L.~Trentadue and G.~Veneziano, {\it {Fracture functions from cut
  vertices}},  \href{http://dx.doi.org/10.1016/S0550-3213(97)00840-7}{{\em
  Nucl. Phys. B} {\bf 519} (1998) 394}
  [\href{http://arXiv.org/abs/hep-ph/9709452}{{\tt arXiv:hep-ph/9709452}}].

\bibitem{Berera:1995fj}
A.~Berera and D.~E. Soper, {\it {Behavior of diffractive parton distribution
  functions}},  \href{http://dx.doi.org/10.1103/PhysRevD.53.6162}{{\em Phys.
  Rev. D} {\bf 53} (1996) 6162}
  [\href{http://arXiv.org/abs/hep-ph/9509239}{{\tt arXiv:hep-ph/9509239}}].

\bibitem{Anselmino:2011ss}
M.~Anselmino, V.~Barone and A.~Kotzinian, {\it {SIDIS in the target
  fragmentation region: Polarized and transverse momentum dependent fracture
  functions}},  \href{http://dx.doi.org/10.1016/j.physletb.2011.03.067}{{\em
  Phys. Lett. B} {\bf 699} (2011) 108}
  [\href{http://arXiv.org/abs/1102.4214}{{\tt arXiv:1102.4214 [hep-ph]}}].

\bibitem{Collins:1997sr}
J.~C. Collins, {\it {Proof of factorization for diffractive hard scattering}},
  \href{http://dx.doi.org/10.1103/PhysRevD.61.019902}{{\em Phys. Rev. D} {\bf
  57} (1998) 3051} [\href{http://arXiv.org/abs/hep-ph/9709499}{{\tt
  arXiv:hep-ph/9709499}}].
\newblock [Erratum: Phys.Rev.D 61, 019902 (2000)].

\bibitem{Chen:2023wsi}
K.~B. Chen, J.~P. Ma and X.~B. Tong, {\it {Twist-3 contributions in
  semi-inclusive DIS in the target fragmentation region}},
  \href{http://dx.doi.org/10.1103/PhysRevD.108.094015}{{\em Phys. Rev. D} {\bf
  108} (2023)~no.~9 094015} [\href{http://arXiv.org/abs/2308.11251}{{\tt
  arXiv:2308.11251 [hep-ph]}}].

\bibitem{Chen:2024brp}
K.-B. Chen, J.-P. Ma and X.-B. Tong, {\it {Gluonic contributions to
  semi-inclusive DIS in the target fragmentation region}},
  \href{http://dx.doi.org/10.1007/JHEP05(2024)298}{{\em JHEP} {\bf 05} (2024)
  298} [\href{http://arXiv.org/abs/2402.15112}{{\tt arXiv:2402.15112
  [hep-ph]}}].

\bibitem{Chen:2021vby}
K.~B. Chen, J.~P. Ma and X.~B. Tong, {\it {Matching of fracture functions for
  SIDIS in target fragmentation region}},
  \href{http://dx.doi.org/10.1007/JHEP11(2021)038}{{\em JHEP} {\bf 11} (2021)
  038} [\href{http://arXiv.org/abs/2108.13582}{{\tt arXiv:2108.13582
  [hep-ph]}}].

\bibitem{Chai:2019ykk}
X.~P. Chai, K.~B. Chen, J.~P. Ma and X.~B. Tong, {\it {Fracture functions in
  different kinematic regions and their factorizations}},
  \href{http://dx.doi.org/10.1007/JHEP10(2019)285}{{\em JHEP} {\bf 10} (2019)
  285} [\href{http://arXiv.org/abs/1903.00809}{{\tt arXiv:1903.00809
  [hep-ph]}}].

\bibitem{Basham:1978bw}
C.~L. Basham, L.~S. Brown, S.~D. Ellis and S.~T. Love, {\it {Energy
  Correlations in electron - Positron Annihilation: Testing QCD}},
  \href{http://dx.doi.org/10.1103/PhysRevLett.41.1585}{{\em Phys. Rev. Lett.}
  {\bf 41} (1978) 1585}.

\bibitem{Basham:1977iq}
C.~L. Basham, L.~S. Brown, S.~D. Ellis and S.~T. Love, {\it {Electron -
  Positron Annihilation Energy Pattern in Quantum Chromodynamics:
  Asymptotically Free Perturbation Theory}},
  \href{http://dx.doi.org/10.1103/PhysRevD.17.2298}{{\em Phys. Rev. D} {\bf 17}
  (1978) 2298}.

\bibitem{Basham:1978zq}
C.~L. Basham, L.~S. Brown, S.~D. Ellis and S.~T. Love, {\it {Energy
  Correlations in electron-Positron Annihilation in Quantum Chromodynamics:
  Asymptotically Free Perturbation Theory}},
  \href{http://dx.doi.org/10.1103/PhysRevD.19.2018}{{\em Phys. Rev. D} {\bf 19}
  (1979) 2018}.

\bibitem{Chen:2020vvp}
H.~Chen, I.~Moult, X.~Zhang and H.~X. Zhu, {\it {Rethinking jets with energy
  correlators: Tracks, resummation, and analytic continuation}},
  \href{http://dx.doi.org/10.1103/PhysRevD.102.054012}{{\em Phys. Rev. D} {\bf
  102} (2020)~no.~5 054012} [\href{http://arXiv.org/abs/2004.11381}{{\tt
  arXiv:2004.11381 [hep-ph]}}].

\bibitem{Li:2021zcf}
Y.~Li, I.~Moult, S.~S. van Velzen, W.~J. Waalewijn and H.~X. Zhu, {\it
  {Extending Precision Perturbative QCD with Track Functions}},
  \href{http://dx.doi.org/10.1103/PhysRevLett.128.182001}{{\em Phys. Rev.
  Lett.} {\bf 128} (2022)~no.~18 182001}
  [\href{http://arXiv.org/abs/2108.01674}{{\tt arXiv:2108.01674 [hep-ph]}}].

\bibitem{Jaarsma:2022kdd}
M.~Jaarsma, Y.~Li, I.~Moult, W.~Waalewijn and H.~X. Zhu, {\it {Renormalization
  group flows for track function moments}},
  \href{http://dx.doi.org/10.1007/JHEP06(2022)139}{{\em JHEP} {\bf 06} (2022)
  139} [\href{http://arXiv.org/abs/2201.05166}{{\tt arXiv:2201.05166
  [hep-ph]}}].

\bibitem{Jaarsma:2023ell}
M.~Jaarsma, Y.~Li, I.~Moult, W.~J. Waalewijn and H.~X. Zhu, {\it {Energy
  correlators on tracks: resummation and non-perturbative effects}},
  \href{http://dx.doi.org/10.1007/JHEP12(2023)087}{{\em JHEP} {\bf 12} (2023)
  087} [\href{http://arXiv.org/abs/2307.15739}{{\tt arXiv:2307.15739
  [hep-ph]}}].

\bibitem{Lee:2023npz}
K.~Lee and I.~Moult, {\it {Energy Correlators Taking Charge}},
  \href{http://arXiv.org/abs/2308.00746}{{\tt arXiv:2308.00746 [hep-ph]}}.

\bibitem{Lee:2023tkr}
K.~Lee and I.~Moult, {\it {Joint Track Functions: Expanding the Space of
  Calculable Correlations at Colliders}},
  \href{http://arXiv.org/abs/2308.01332}{{\tt arXiv:2308.01332 [hep-ph]}}.

\bibitem{Lee:2022ige}
K.~Lee, B.~Me\c{c}aj and I.~Moult, {\it {Conformal collider physics meets LHC
  data}},  \href{http://dx.doi.org/10.1103/PhysRevD.111.L011502}{{\em Phys.
  Rev. D} {\bf 111} (2025)~no.~1 L011502}
  [\href{http://arXiv.org/abs/2205.03414}{{\tt arXiv:2205.03414 [hep-ph]}}].

\bibitem{Craft:2022kdo}
E.~Craft, K.~Lee, B.~Me\c{c}aj and I.~Moult, {\it {Beautiful and Charming
  Energy Correlators}},  \href{http://arXiv.org/abs/2210.09311}{{\tt
  arXiv:2210.09311 [hep-ph]}}.

\bibitem{Komiske:2022enw}
P.~T. Komiske, I.~Moult, J.~Thaler and H.~X. Zhu, {\it {Analyzing N-Point
  Energy Correlators inside Jets with CMS Open Data}},
  \href{http://dx.doi.org/10.1103/PhysRevLett.130.051901}{{\em Phys. Rev.
  Lett.} {\bf 130} (2023)~no.~5 051901}
  [\href{http://arXiv.org/abs/2201.07800}{{\tt arXiv:2201.07800 [hep-ph]}}].

\bibitem{CMS:2024mlf}
{\bf CMS} collaboration, A.~Hayrapetyan {\em et.~al.}, {\it {Measurement of
  Energy Correlators inside Jets and Determination of the Strong Coupling
  \ensuremath{\alpha}S(mZ)}},
  \href{http://dx.doi.org/10.1103/PhysRevLett.133.071903}{{\em Phys. Rev.
  Lett.} {\bf 133} (2024)~no.~7 071903}
  [\href{http://arXiv.org/abs/2402.13864}{{\tt arXiv:2402.13864 [hep-ex]}}].

\bibitem{ALICE:2024dfl}
{\bf ALICE} collaboration, S.~Acharya {\em et.~al.}, {\it {Exposing the
  parton-hadron transition within jets with energy-energy correlators in pp
  collisions at $\sqrt{\textit s}=5.02$ TeV}},
  \href{http://arXiv.org/abs/2409.12687}{{\tt arXiv:2409.12687 [hep-ex]}}.

\bibitem{Liu:2024lxy}
X.~Liu, W.~Vogelsang, F.~Yuan and H.~X. Zhu, {\it {Universality in the
  Near-Side Energy-Energy Correlator}},
  \href{http://arXiv.org/abs/2410.16371}{{\tt arXiv:2410.16371 [hep-ph]}}.

\bibitem{Lee:2024esz}
K.~Lee, A.~Pathak, I.~W. Stewart and Z.~Sun, {\it {Nonperturbative Effects in
  Energy Correlators: From Characterizing Confinement Transition to Improving
  \ensuremath{\alpha}s Extraction}},
  \href{http://dx.doi.org/10.1103/PhysRevLett.133.231902}{{\em Phys. Rev.
  Lett.} {\bf 133} (2024)~no.~23 231902}
  [\href{http://arXiv.org/abs/2405.19396}{{\tt arXiv:2405.19396 [hep-ph]}}].

\bibitem{Barata:2024wsu}
J.~a. Barata, Z.-B. Kang, X.~Mayo~L\'opez and J.~Penttala, {\it {Energy-Energy
  Correlator for jet production in $pp$ and $pA$ collisions}},
  \href{http://arXiv.org/abs/2411.11782}{{\tt arXiv:2411.11782 [hep-ph]}}.

\bibitem{Alipour-fard:2024szj}
S.~Alipour-fard, A.~Budhraja, J.~Thaler and W.~J. Waalewijn, {\it {New Angles
  on Energy Correlators}},  \href{http://arXiv.org/abs/2410.16368}{{\tt
  arXiv:2410.16368 [hep-ph]}}.

\bibitem{Alipour-fard:2025dvp}
S.~Alipour-fard and W.~J. Waalewijn, {\it {Energy Correlators Beyond Angles}},
  \href{http://arXiv.org/abs/2501.17218}{{\tt arXiv:2501.17218 [hep-ph]}}.

\bibitem{Andres:2022ovj}
C.~Andres, F.~Dominguez, R.~Kunnawalkam~Elayavalli, J.~Holguin, C.~Marquet and
  I.~Moult, {\it {Resolving the Scales of the Quark-Gluon Plasma with Energy
  Correlators}},  \href{http://dx.doi.org/10.1103/PhysRevLett.130.262301}{{\em
  Phys. Rev. Lett.} {\bf 130} (2023)~no.~26 262301}
  [\href{http://arXiv.org/abs/2209.11236}{{\tt arXiv:2209.11236 [hep-ph]}}].

\bibitem{Andres:2023xwr}
C.~Andres, F.~Dominguez, J.~Holguin, C.~Marquet and I.~Moult, {\it {A coherent
  view of the quark-gluon plasma from energy correlators}},
  \href{http://dx.doi.org/10.1007/JHEP09(2023)088}{{\em JHEP} {\bf 09} (2023)
  088} [\href{http://arXiv.org/abs/2303.03413}{{\tt arXiv:2303.03413
  [hep-ph]}}].

\bibitem{Devereaux:2023vjz}
K.~Devereaux, W.~Fan, W.~Ke, K.~Lee and I.~Moult, {\it {Imaging Cold Nuclear
  Matter with Energy Correlators}},
  \href{http://arXiv.org/abs/2303.08143}{{\tt arXiv:2303.08143 [hep-ph]}}.

\bibitem{Andres:2023ymw}
C.~Andres, F.~Dominguez, J.~Holguin, C.~Marquet and I.~Moult, {\it {Seeing
  beauty in the quark-gluon plasma with energy correlators}},
  \href{http://dx.doi.org/10.1103/PhysRevD.110.L031503}{{\em Phys. Rev. D} {\bf
  110} (2024)~no.~3 L031503} [\href{http://arXiv.org/abs/2307.15110}{{\tt
  arXiv:2307.15110 [hep-ph]}}].

\bibitem{Yang:2023dwc}
Z.~Yang, Y.~He, I.~Moult and X.-N. Wang, {\it {Probing the Short-Distance
  Structure of the Quark-Gluon Plasma with Energy Correlators}},
  \href{http://dx.doi.org/10.1103/PhysRevLett.132.011901}{{\em Phys. Rev.
  Lett.} {\bf 132} (2024)~no.~1 011901}
  [\href{http://arXiv.org/abs/2310.01500}{{\tt arXiv:2310.01500 [hep-ph]}}].

\bibitem{Barata:2023bhh}
J.~a. Barata, P.~Caucal, A.~Soto-Ontoso and R.~Szafron, {\it {Advancing the
  understanding of energy-energy correlators in heavy-ion collisions}},
  \href{http://dx.doi.org/10.1007/JHEP11(2024)060}{{\em JHEP} {\bf 11} (2024)
  060} [\href{http://arXiv.org/abs/2312.12527}{{\tt arXiv:2312.12527
  [hep-ph]}}].

\bibitem{Barata:2023zqg}
J.~a. Barata, J.~G. Milhano and A.~V. Sadofyev, {\it {Picturing QCD jets in
  anisotropic matter: from jet shapes to energy energy correlators}},
  \href{http://dx.doi.org/10.1140/epjc/s10052-024-12514-1}{{\em Eur. Phys. J.
  C} {\bf 84} (2024)~no.~2 174} [\href{http://arXiv.org/abs/2308.01294}{{\tt
  arXiv:2308.01294 [hep-ph]}}].

\bibitem{Bossi:2024qho}
H.~Bossi, A.~S. Kudinoor, I.~Moult, D.~Pablos, A.~Rai and K.~Rajagopal, {\it
  {Imaging the wakes of jets with energy-energy-energy correlators}},
  \href{http://dx.doi.org/10.1007/JHEP12(2024)073}{{\em JHEP} {\bf 12} (2024)
  073} [\href{http://arXiv.org/abs/2407.13818}{{\tt arXiv:2407.13818
  [hep-ph]}}].

\bibitem{Xing:2024yrb}
W.-J. Xing, S.~Cao, G.-Y. Qin and X.-N. Wang, {\it {Flavor Hierarchy of Jet
  Energy Correlators inside the Quark-Gluon Plasma}},
  \href{http://dx.doi.org/10.1103/PhysRevLett.134.052301}{{\em Phys. Rev.
  Lett.} {\bf 134} (2025)~no.~5 052301}
  [\href{http://arXiv.org/abs/2409.12843}{{\tt arXiv:2409.12843 [hep-ph]}}].

\bibitem{Fu:2024pic}
Y.~Fu, B.~M\"uller and C.~Sirimanna, {\it {Modification of the Jet
  Energy-Energy Correlator in Cold Nuclear Matter}},
  \href{http://arXiv.org/abs/2411.04866}{{\tt arXiv:2411.04866 [nucl-th]}}.

\bibitem{Andres:2024xvk}
C.~Andres, F.~Dominguez, J.~Holguin, C.~Marquet and I.~Moult, {\it {Simple
  Scaling Laws for Energy Correlators in Nuclear Matter}},
  \href{http://arXiv.org/abs/2411.15298}{{\tt arXiv:2411.15298 [hep-ph]}}.

\bibitem{Barata:2025fzd}
J.~a. Barata, I.~Moult, A.~V. Sadofyev and J.~a.~M. Silva, {\it {Dissecting Jet
  Modification in the QGP with Multi-Point Energy Correlators}},
  \href{http://arXiv.org/abs/2503.13603}{{\tt arXiv:2503.13603 [hep-ph]}}.

\bibitem{Apolinario:2025vtx}
L.~Apolin\'ario, R.~Kunnawalkam~Elayavalli, N.~O. Madureira, J.-X. Sheng, X.-N.
  Wang and Z.~Yang, {\it {Flavor dependence of Energy-energy correlators}},
  \href{http://arXiv.org/abs/2502.11406}{{\tt arXiv:2502.11406 [hep-ph]}}.

\bibitem{Meng:1991da}
R.-b. Meng, F.~I. Olness and D.~E. Soper, {\it {Semiinclusive deeply inelastic
  scattering at electron - proton colliders}},
  \href{http://dx.doi.org/10.1016/0550-3213(92)90230-9}{{\em Nucl. Phys. B}
  {\bf 371} (1992) 79}.

\bibitem{Li:2021txc}
H.~T. Li, Y.~Makris and I.~Vitev, {\it {Energy-energy correlators in Deep
  Inelastic Scattering}},
  \href{http://dx.doi.org/10.1103/PhysRevD.103.094005}{{\em Phys. Rev. D} {\bf
  103} (2021)~no.~9 094005} [\href{http://arXiv.org/abs/2102.05669}{{\tt
  arXiv:2102.05669 [hep-ph]}}].

\bibitem{Kang:2023big}
Z.-B. Kang, K.~Lee, D.~Y. Shao and F.~Zhao, {\it {Probing transverse momentum
  dependent structures with azimuthal dependence of energy correlators}},
  \href{http://dx.doi.org/10.1007/JHEP03(2024)153}{{\em JHEP} {\bf 03} (2024)
  153} [\href{http://arXiv.org/abs/2310.15159}{{\tt arXiv:2310.15159
  [hep-ph]}}].

\bibitem{Liu:2022wop}
X.~Liu and H.~X. Zhu, {\it {Nucleon Energy Correlators}},
  \href{http://dx.doi.org/10.1103/PhysRevLett.130.091901}{{\em Phys. Rev.
  Lett.} {\bf 130} (2023)~no.~9 091901}
  [\href{http://arXiv.org/abs/2209.02080}{{\tt arXiv:2209.02080 [hep-ph]}}].

\bibitem{Chen:2024bpj}
K.-B. Chen, J.-P. Ma and X.-B. Tong, {\it {The connection between nucleon
  energy correlators and fracture functions}},
  \href{http://dx.doi.org/10.1007/JHEP08(2024)227}{{\em JHEP} {\bf 08} (2024)
  227} [\href{http://arXiv.org/abs/2406.08559}{{\tt arXiv:2406.08559
  [hep-ph]}}].

\bibitem{Liu:2024kqt}
X.~Liu and H.~X. Zhu, {\it {TMDs from Semi-inclusive Energy Correlators}},
  \href{http://arXiv.org/abs/2403.08874}{{\tt arXiv:2403.08874 [hep-ph]}}.

\bibitem{Cao:2023oef}
H.~Cao, X.~Liu and H.~X. Zhu, {\it {Toward precision measurements of nucleon
  energy correlators in lepton-nucleon collisions}},
  \href{http://dx.doi.org/10.1103/PhysRevD.107.114008}{{\em Phys. Rev. D} {\bf
  107} (2023)~no.~11 114008} [\href{http://arXiv.org/abs/2303.01530}{{\tt
  arXiv:2303.01530 [hep-ph]}}].

\bibitem{Liu:2023aqb}
H.-Y. Liu, X.~Liu, J.-C. Pan, F.~Yuan and H.~X. Zhu, {\it {Nucleon Energy
  Correlators for the Color Glass Condensate}},
  \href{http://dx.doi.org/10.1103/PhysRevLett.130.181901}{{\em Phys. Rev.
  Lett.} {\bf 130} (2023)~no.~18 181901}
  [\href{http://arXiv.org/abs/2301.01788}{{\tt arXiv:2301.01788 [hep-ph]}}].

\bibitem{Li:2023gkh}
X.~L. Li, X.~Liu, F.~Yuan and H.~X. Zhu, {\it {Illuminating nucleon-gluon
  interference via calorimetric asymmetry}},
  \href{http://dx.doi.org/10.1103/PhysRevD.108.L091502}{{\em Phys. Rev. D} {\bf
  108} (2023)~no.~9 L091502} [\href{http://arXiv.org/abs/2308.10942}{{\tt
  arXiv:2308.10942 [hep-ph]}}].

\bibitem{Guo:2024jch}
Y.~Guo, X.~Liu, F.~Yuan and H.~X. Zhu, {\it {Long-Range Azimuthal Correlation,
  Entanglement, and Bell Inequality Violation by Spinning Gluons at the Large
  Hadron Collider}},  \href{http://dx.doi.org/10.34133/research.0552}{{\em
  Research} {\bf 2025} (2025) 0552}
  [\href{http://arXiv.org/abs/2406.05880}{{\tt arXiv:2406.05880 [hep-ph]}}].

\bibitem{Guo:2024vpe}
Y.~Guo, X.~Liu and F.~Yuan, {\it {Long Range Energy-energy Correlator at the
  LHC}},  \href{http://arXiv.org/abs/2408.14693}{{\tt arXiv:2408.14693
  [hep-ph]}}.

\bibitem{Cao:2023qat}
H.~Cao, H.~T. Li and Z.~Mi, {\it {Bjorken x weighted energy-energy correlators
  from the target fragmentation region to the current fragmentation region}},
  \href{http://dx.doi.org/10.1103/PhysRevD.109.096004}{{\em Phys. Rev. D} {\bf
  109} (2024)~no.~9 096004} [\href{http://arXiv.org/abs/2312.07655}{{\tt
  arXiv:2312.07655 [hep-ph]}}].

\bibitem{Caucal:2025qjg}
P.~Caucal and F.~Salazar, {\it {Transverse momentum dependent factorisation in
  the target fragmentation region at small $x$}},
  \href{http://arXiv.org/abs/2502.02634}{{\tt arXiv:2502.02634 [hep-ph]}}.

\bibitem{Sveshnikov:1995vi}
N.~A. Sveshnikov and F.~V. Tkachov, {\it {Jets and quantum field theory}},
  \href{http://dx.doi.org/10.1016/0370-2693(96)00558-8}{{\em Phys. Lett. B}
  {\bf 382} (1996) 403} [\href{http://arXiv.org/abs/hep-ph/9512370}{{\tt
  arXiv:hep-ph/9512370}}].

\bibitem{Bauer:2008dt}
C.~W. Bauer, S.~P. Fleming, C.~Lee and G.~F. Sterman, {\it {Factorization of
  $e^+e^-$ Event Shape Distributions with Hadronic Final States in Soft
  Collinear Effective Theory}},
  \href{http://dx.doi.org/10.1103/PhysRevD.78.034027}{{\em Phys. Rev. D} {\bf
  78} (2008) 034027} [\href{http://arXiv.org/abs/0801.4569}{{\tt
  arXiv:0801.4569 [hep-ph]}}].

\bibitem{Anselmino:2011bb}
M.~Anselmino, V.~Barone and A.~Kotzinian, {\it {Double hadron lepto-production
  in the current and target fragmentation regions}},
  \href{http://dx.doi.org/10.1016/j.physletb.2011.10.064}{{\em Phys. Lett. B}
  {\bf 706} (2011) 46} [\href{http://arXiv.org/abs/1109.1132}{{\tt
  arXiv:1109.1132 [hep-ph]}}].

\bibitem{Anselmino:2011vkz}
M.~Anselmino, V.~Barone and A.~Kotzinian, {\it {A novel beam-spin asymmetry in
  double-hadron inclusive lepto-production}},
  \href{http://dx.doi.org/10.1016/j.physletb.2012.06.003}{{\em Phys. Lett. B}
  {\bf 713} (2012) 317} [\href{http://arXiv.org/abs/1112.2604}{{\tt
  arXiv:1112.2604 [hep-ph]}}].

\bibitem{CLAS:2022sqt}
{\bf CLAS} collaboration, H.~Avakian {\em et.~al.}, {\it {Observation of
  Correlations between Spin and Transverse Momenta in Back-to-Back Dihadron
  Production at CLAS12}},
  \href{http://dx.doi.org/10.1103/PhysRevLett.130.022501}{{\em Phys. Rev.
  Lett.} {\bf 130} (2023)~no.~2 022501}
  [\href{http://arXiv.org/abs/2208.05086}{{\tt arXiv:2208.05086 [hep-ex]}}].

\bibitem{Guo:2023uis}
Y.~Guo and F.~Yuan, {\it {Explore the Nucleon Tomography through Di-hadron
  Correlation in Opposite Hemisphere in Deep Inelastic Scattering}},
  \href{http://arXiv.org/abs/2312.01008}{{\tt arXiv:2312.01008 [hep-ph]}}.

\bibitem{Zhao:2024usu}
X.~Zhao, Z.-t. Liang, T.~Liu and Y.-j. Zhou, {\it {Suppression of Spin Transfer
  to $\Lambda$ in Deep Inelastic Scattering}},
  \href{http://arXiv.org/abs/2411.06205}{{\tt arXiv:2411.06205 [hep-ph]}}.

\bibitem{Gelis:2010nm}
F.~Gelis, E.~Iancu, J.~Jalilian-Marian and R.~Venugopalan, {\it {The Color
  Glass Condensate}},
  \href{http://dx.doi.org/10.1146/annurev.nucl.010909.083629}{{\em Ann. Rev.
  Nucl. Part. Sci.} {\bf 60} (2010) 463}
  [\href{http://arXiv.org/abs/1002.0333}{{\tt arXiv:1002.0333 [hep-ph]}}].

\bibitem{Iancu:2003xm}
E.~Iancu and R.~Venugopalan, {\em {The Color glass condensate and high-energy
  scattering in QCD}}, pp.~249--3363.
\newblock 3, 2003.
\newblock \href{http://arXiv.org/abs/hep-ph/0303204}{{\tt
  arXiv:hep-ph/0303204}}.

\bibitem{Marquet:2009ca}
C.~Marquet, B.-W. Xiao and F.~Yuan, {\it {Semi-inclusive Deep Inelastic
  Scattering at small x}},
  \href{http://dx.doi.org/10.1016/j.physletb.2009.10.099}{{\em Phys. Lett. B}
  {\bf 682} (2009) 207} [\href{http://arXiv.org/abs/0906.1454}{{\tt
  arXiv:0906.1454 [hep-ph]}}].

\bibitem{Xiao:2017yya}
B.-W. Xiao, F.~Yuan and J.~Zhou, {\it {Transverse Momentum Dependent Parton
  Distributions at Small-x}},
  \href{http://dx.doi.org/10.1016/j.nuclphysb.2017.05.012}{{\em Nucl. Phys. B}
  {\bf 921} (2017) 104} [\href{http://arXiv.org/abs/1703.06163}{{\tt
  arXiv:1703.06163 [hep-ph]}}].

\bibitem{Chang:2013rca}
H.-M. Chang, M.~Procura, J.~Thaler and W.~J. Waalewijn, {\it {Calculating
  Track-Based Observables for the LHC}},
  \href{http://dx.doi.org/10.1103/PhysRevLett.111.102002}{{\em Phys. Rev.
  Lett.} {\bf 111} (2013) 102002} [\href{http://arXiv.org/abs/1303.6637}{{\tt
  arXiv:1303.6637 [hep-ph]}}].

\bibitem{Chang:2013iba}
H.-M. Chang, M.~Procura, J.~Thaler and W.~J. Waalewijn, {\it {Calculating Track
  Thrust with Track Functions}},
  \href{http://dx.doi.org/10.1103/PhysRevD.88.034030}{{\em Phys. Rev. D} {\bf
  88} (2013) 034030} [\href{http://arXiv.org/abs/1306.6630}{{\tt
  arXiv:1306.6630 [hep-ph]}}].

\bibitem{Bertone:2018ecm}
{\bf NNPDF} collaboration, V.~Bertone, N.~P. Hartland, E.~R. Nocera, J.~Rojo
  and L.~Rottoli, {\it {Charged hadron fragmentation functions from collider
  data}},  \href{http://dx.doi.org/10.1140/epjc/s10052-018-6130-4}{{\em Eur.
  Phys. J. C} {\bf 78} (2018)~no.~8 651}
  [\href{http://arXiv.org/abs/1807.03310}{{\tt arXiv:1807.03310 [hep-ph]}}].
\newblock [Erratum: Eur.Phys.J.C 84, 155 (2024)].

\bibitem{Bertone:2017tyb}
{\bf NNPDF} collaboration, V.~Bertone, S.~Carrazza, N.~P. Hartland, E.~R.
  Nocera and J.~Rojo, {\it {A determination of the fragmentation functions of
  pions, kaons, and protons with faithful uncertainties}},
  \href{http://dx.doi.org/10.1140/epjc/s10052-017-5088-y}{{\em Eur. Phys. J. C}
  {\bf 77} (2017)~no.~8 516} [\href{http://arXiv.org/abs/1706.07049}{{\tt
  arXiv:1706.07049 [hep-ph]}}].

\bibitem{Golec-Biernat:1998zce}
K.~J. Golec-Biernat and M.~Wusthoff, {\it {Saturation effects in deep inelastic
  scattering at low Q**2 and its implications on diffraction}},
  \href{http://dx.doi.org/10.1103/PhysRevD.59.014017}{{\em Phys. Rev. D} {\bf
  59} (1998) 014017} [\href{http://arXiv.org/abs/hep-ph/9807513}{{\tt
  arXiv:hep-ph/9807513}}].

\bibitem{Lappi:2013zma}
T.~Lappi and H.~M\"antysaari, {\it {Single inclusive particle production at
  high energy from HERA data to proton-nucleus collisions}},
  \href{http://dx.doi.org/10.1103/PhysRevD.88.114020}{{\em Phys. Rev. D} {\bf
  88} (2013) 114020} [\href{http://arXiv.org/abs/1309.6963}{{\tt
  arXiv:1309.6963 [hep-ph]}}].

\bibitem{Golec-Biernat:2017lfv}
K.~Golec-Biernat and S.~Sapeta, {\it {Saturation model of DIS : an update}},
  \href{http://dx.doi.org/10.1007/JHEP03(2018)102}{{\em JHEP} {\bf 03} (2018)
  102} [\href{http://arXiv.org/abs/1711.11360}{{\tt arXiv:1711.11360
  [hep-ph]}}].

\bibitem{Kovchegov:1999yj}
Y.~V. Kovchegov, {\it {Small x F(2) structure function of a nucleus including
  multiple pomeron exchanges}},
  \href{http://dx.doi.org/10.1103/PhysRevD.60.034008}{{\em Phys. Rev. D} {\bf
  60} (1999) 034008} [\href{http://arXiv.org/abs/hep-ph/9901281}{{\tt
  arXiv:hep-ph/9901281}}].

\bibitem{Balitsky:1995ub}
I.~Balitsky, {\it {Operator expansion for high-energy scattering}},
  \href{http://dx.doi.org/10.1016/0550-3213(95)00638-9}{{\em Nucl. Phys. B}
  {\bf 463} (1996) 99} [\href{http://arXiv.org/abs/hep-ph/9509348}{{\tt
  arXiv:hep-ph/9509348}}].

\bibitem{Balitsky:2006wa}
I.~Balitsky, {\it {Quark contribution to the small-x evolution of color
  dipole}},  \href{http://dx.doi.org/10.1103/PhysRevD.75.014001}{{\em Phys.
  Rev. D} {\bf 75} (2007) 014001}
  [\href{http://arXiv.org/abs/hep-ph/0609105}{{\tt arXiv:hep-ph/0609105}}].

\bibitem{DAlesio:2015fwo}
U.~D'Alesio, F.~Murgia and C.~Pisano, {\it {Towards a first estimate of the
  gluon Sivers function from A$_{N}$ data in pp collisions at RHIC}},
  \href{http://dx.doi.org/10.1007/JHEP09(2015)119}{{\em JHEP} {\bf 09} (2015)
  119} [\href{http://arXiv.org/abs/1506.03078}{{\tt arXiv:1506.03078
  [hep-ph]}}].

\bibitem{Gluck:1998xa}
M.~Gl\"uck, E.~Reya and A.~Vogt, {\it {Dynamical parton distributions
  revisited}},  \href{http://dx.doi.org/10.1007/s100520050289}{{\em Eur. Phys.
  J. C} {\bf 5} (1998) 461} [\href{http://arXiv.org/abs/hep-ph/9806404}{{\tt
  arXiv:hep-ph/9806404}}].

\bibitem{Santiago:2024iem}
M.~G. Santiago, D.~Adamiak and Y.~Tawabutr, {\it {Leading-Twist Flavor Singlet
  Quark TMDs at Small-$x$}},  \href{http://arXiv.org/abs/2412.14154}{{\tt
  arXiv:2412.14154 [hep-ph]}}.

\bibitem{Schafer:2013opa}
A.~Sch\"afer and J.~Zhou, {\it {A note on the scale evolution of tri-gluon
  correlations}},  \href{http://arXiv.org/abs/1308.4961}{{\tt arXiv:1308.4961
  [hep-ph]}}.

\end{thebibliography}\endgroup

\end{document}